\documentclass[acmsmall]{acmart}

\AtBeginDocument{%
  \providecommand\BibTeX{{%
    \normalfont B\kern-0.5em{\scshape i\kern-0.25em b}\kern-0.8em\TeX}}}

\setcopyright{acmcopyright}

\acmJournal{TOSN}

\usepackage{hyperref}
\usepackage{url}
\usepackage{multirow}
\usepackage[english]{babel}
\usepackage[utf8]{inputenc}
\usepackage{algorithm} \usepackage{algorithmicx}
\usepackage{algpseudocode}
\usepackage{enumerate}
\usepackage{enumitem}
\usepackage{graphics}
\usepackage{graphicx}
\usepackage{amsmath}
\usepackage{ragged2e}
\usepackage{threeparttable} 
\usepackage[caption=false,font=normalsize]{subfig}
\usepackage{xspace}
\usepackage{makecell}
\usepackage{bbding}
\usepackage{verbatim}
\usepackage{color}
\usepackage{xcolor}

\newcommand{\ie}{\emph{i.e.},\xspace}

\newcommand{\eg}{\emph{e.g.},\xspace}

\newcommand{\systemname}{{AdaMEC}\xspace}
\newcommand{\systemnameposs}{{AdaMEC's}\xspace}

\begin{document}

\title{AdaMEC: Towards a Context-Adaptive and Dynamically-Combinable DNN Deployment Framework for Mobile Edge Computing}


\author{BoWen Pang}
\thanks{This work was partially supported by the National Key R\&D Program of China (2019YFB1703901), the National Science Fund for Distinguished Young Scholars (62025205),  and the National Natural Science Foundation of China (No. 62032020, 61725205, 62102317).}
\affiliation{%
  \institution{Northwestern Polytechnical University}
  \streetaddress{No. 127, Youyi-West Rd}
  \city{Xi’an}
  \state{Shaanxi}
  \country{China}
  \postcode{710072}
}

\author{Sicong Liu}
\affiliation{%
  \institution{Northwestern Polytechnical University}
  \streetaddress{No. 127, Youyi-West Rd}
  \city{Xi’an}
  \state{Shaanxi}
  \country{China}
  \postcode{710072}
}

\author{Hongli Wang}
\affiliation{%
  \institution{Northwestern Polytechnical University}
  \streetaddress{No. 127, Youyi-West Rd}
  \city{Xi’an}
  \state{Shaanxi}
  \country{China}
  \postcode{710072}
}

\author{Bin Guo}
\authornote{Corresponding author}
\email{guob@nwpu.edu.cn (Corresponding author)}
\affiliation{%
  \institution{Northwestern Polytechnical University}
  \streetaddress{No. 127, Youyi-West Rd}
  \city{Xi’an}
  \state{Shaanxi}
  \country{China}
  \postcode{710072}
}

\author{Yuzhan Wang}
\affiliation{%
  \institution{Northwestern Polytechnical University}
  \streetaddress{No. 127, Youyi-West Rd}
  \city{Xi’an}
  \state{Shaanxi}
  \country{China}
  \postcode{710072}
}

\author{Hao Wang}
\affiliation{%
  \institution{Northwestern Polytechnical University}
  \streetaddress{No. 127, Youyi-West Rd}
  \city{Xi’an}
  \state{Shaanxi}
  \country{China}
  \postcode{710072}
}

\author{Zhenli Sheng}
\affiliation{%
 \institution{Huawei Technologies}
 \city{Hangzhou}
 \state{Zhejiang}
 \country{China}
}

\author{Zhongyi Wang}
\affiliation{%
  \institution{Huawei Technologies}
  \city{Hangzhou}
  \state{Zhejiang}
  \country{China}
}

\author{Zhiwen Yu}
\affiliation{%
 \institution{Northwestern Polytechnical University}
  \streetaddress{No. 127, Youyi-West Rd}
  \city{Xi’an}
  \state{Shaanxi}
  \country{China}
  \postcode{710072}
}

\renewcommand{\shortauthors}{Wang, et al.}


\begin{abstract}
\justifying
  With the rapid development of deep learning, recent research on intelligent and interactive mobile applications (\eg health monitoring, speech recognition) has attracted extensive attention. 
  And these applications necessitate the mobile edge computing scheme, \ie offloading partial computation from mobile devices to edge devices for inference acceleration and transmission load reduction. The current practices have relied on collaborative DNN partition and offloading to satisfy the predefined latency requirements, which is intractable to adapt to the dynamic deployment context at runtime.
  \systemname, a context-adaptive and dynamically-combinable DNN deployment framework is proposed to meet these requirements for mobile edge computing, which consists of three novel techniques.
  First, \textit{once-for-all DNN pre-partition} divides DNN at the primitive operator level and stores partitioned modules into executable files, defined as pre-partitioned DNN atoms.
  Second, \textit{context-adaptive DNN atom combination and offloading} introduces a graph-based decision algorithm to quickly search the suitable combination of atoms and adaptively make the offloading plan under dynamic deployment contexts.
  Third, \textit{runtime latency predictor} provides timely latency feedback for DNN deployment considering both DNN configurations and dynamic contexts.
  Extensive experiments demonstrate that \systemname outperforms state-of-the-art baselines in terms of latency reduction by up to 62.14\% and average memory saving by 55.21\%.
\end{abstract}

\ccsdesc[500]{Human-centered computing~Ubiquitous computing}
\ccsdesc[300]{Human-centered computing~Ubiquitous and mobile computing systems and tools}
\ccsdesc[100]{Computing methodologies~Cooperation and coordination}

\keywords{Context-adaptive, DNN combination and offloading, DNN partition, edge intelligence}



\maketitle
\section{Introduction}
\label{sec:introduction}

Deep Neural Networks (DNNs) have stimulated a wide range of intelligent applications on mobile and embedded devices, 
such as UbiEar~\cite{sicong2017ubiear} (the acoustic event sensing and notification system on smartphone), Smart-U~\cite{huang2018smart} (the utensil-based meal composition tracking and detection equipment), and DeepEye~\cite{mathur2017deepeye} (the mobile vision system running on wearables). 
However, deploying compute-intensive DNNs (\eg GoogLeNet ~\cite{szegedy2015googlenet}, ResNet~\cite{li2016resnet101}, and BERT ~\cite{kenton2019bert}) on mobile devices is challenging to simultaneously satisfy the application demands on accuracy, resource, and latency.
The reasons are two-fold.
\textit{First}, mobile applications are usually interactive with users and sensitive to inference latency, which requires sufficient computing resources. 
Nevertheless, mobile device resources (\eg memory and processor) always dynamically change.
\textit{Second}, it's prohibitive to deploy the full-size DNN on resource-constrained mobile devices directly. 
While most of the DNN compression techniques (\eg pruning~\cite{he2018multi,gao2021pruning}, decomposition~\cite{wu2018decomposition,liu2021adaspring}, and quantization~\cite{zhu2018quantization,zhang2018lq}) come at the cost of accuracy loss and have the limited ability to accelerate DNN inference~\cite{deng2020compression}.

Consequently, researchers presented computation offloading approaches to aggregate resources of multiple devices, thus improving the overall computing efficiency~\cite{mach2017offloadsurvey,xue2021ddpqn,xue2021eosdnn}.
They mainly contain two categories, \ie offloading partial complex computation to the cloud or edge servers.
For example, Lee et al.,~\cite{lee2017real} offloads the computation to an off-board computing cloud, enabling real-time object detection on Unmanned Aerial Vehicles.
And Hossain~\cite{hossain2019emotion} proposes an edge-cloud-based emotion recognition system in which IoT devices capture facial images and speech signals before distributing them to different edges.
This is in line with the trends on \textit{mobile edge computing (MEC)}, \ie bringing partial computation closer to the data with distributed compute resources across networks~\cite{mao2017mec}. MEC's potential has been demonstrated in various emerging 5G use cases, such as autonomous vehicle functionality, intelligent manufacturing, and disaster monitoring. 
However, these works assume that mobile devices can establish stable connections to edge servers, which is unrealistic due to the \textit{mobility} of mobile devices.

Therefore, the context-adaptive deployment of DNNs is required to perform collaborative computing with mobile edge devices (\eg wearables, smartphones, and IoT facilities) with heterogeneous and dynamic resource availability.
To this end, the DNN partition technique~\cite{kang2017neurosurgeon,hu2019dads,wang2021context} becomes an ideal solution to deploy DNNs on heterogeneous mobile and edge clusters.
Existing efforts explore partitioning a DNN at several points and computing partial modules on different edge devices in series~\cite{kang2017neurosurgeon, laskaridis2020spinn} (partitioned between DNN layers) or parallel~\cite{zhao2018deepthings, mao2017modnn} (partitioned within layers). 
To cope with the dynamic running context, they  need to search for new appropriate partition points from scatch once the running context changes.
Here, from a system perspective, we distinguish between the \textit{end-to-end DNN deployment process} for mobile edge computing and the individual \textit{DNN partition techniques}.
Existing DNN partition practices lack consideration of the practical deployment process of DNNs, thus still facing two technical challenges to satisfy the requirements of latency-sensitive mobile applications. 
\begin{itemize}
    \item The end-to-end deployment process will be disrupted by several difficulties in the real-world MEC deployment context with \textit{dynamically changing} resource availability, network conditions, and latency requirements.
    Any changes in the context will affect the whole deployment process, including DNN re-partition and re-compilation, storage, re-transmission, and re-offloading of partitioned modules, which is costly and intractable. The critical problem is the coupling between DNN partition and offloading.
    \item The heterogeneity of edge devices and the dynamics of network conditions bring great deployment decision space and difficulty in latency estimation. 
It is non-trivial to provide timely and accurate latency feedback for selecting the most suitable DNN deployment plan.
\end{itemize}

To address above challenges, we propose \systemname, a context-adaptive and dynamically-combinable DNN deployment framework for mobile edge computing.
It decouples the end-to-end DNN deployment process into \textit{DNN pre-partition}, and \textit{DNN atom combination and offloading}.
%
%
Specifically, to be independent of deployment requirements and support the efficient collaborative computing in the \textit{arbitrary} and \textit{dynamic} context, the \textit{once-for-all DNN pre-partition} block pre-partitions the DNN model at the most fine-grained level (\eg primitive operators) and filters the candidate partition points based on a latency benefit function.
These pre-partitioned DNN modules are defined as \textit{pre-partitioned DNN atoms}.
Then the \textit{context-adaptive DNN atom combination and offloading} block establishes a search graph and utilizes an adaptive decision algorithm to select the most suitable combination of pre-partitioned DNN atoms efficiently.
Then these atoms are offloaded into corresponding devices to achieve the DNN deployment process.
%
In addition, \systemname adopts the \textit{runtime latency prediction} block to provide the fine-grained partition module-based runtime latency predictor for the above two blocks.
We have implemented \systemname on multiple mobile and edge platforms (\eg Smart Watch, RaspberryPi 4B, and Jetson AGX Xavier) in PyTorch, with CIFAR100~\cite{krizhevsky2009Cifar100}, ImageNet~\cite{deng2009imagenet}, and BDD100K~\cite{yu2018bdd100k} datasets. And it achieves a significant improvement in computation offloading acceleration and memory saving. 

%
The main contributions are as follows.
\begin{itemize} 
    \item To the best of our knowledge, this is the first to present the context-adaptive and dynamically-combinable DNN deployment problem by making a distinction between the \textit{end-to-end DNN deployment} at mobile and edge devices and the original \textit{DNN partition}.
    \item \systemname integrates a series of novel mechanisms (\ie the once-for-all DNN pre-partition at the operator-level and the context-adaptive DNN atom offloading) to boost the efficiency of adaptive DNN deployment under the practical and dynamic deployment context. Also, we present a novel runtime latency predictor to provide timely feedback for \systemname.
    \item Experiments show that \systemname achieves significant memory saving for deploying DNNs on devices (\eg the average 44.56\% and 55.21\% savings compared with baselines for GoogLeNet under dynamic mobile edge computing contexts). 
    It remarkably accelerates the computation offloading process, enabling the mobile device timely obtain the offloading benefits (\eg 35.04\% $\sim$ 62.14\% faster than baselines). 
\end{itemize}
\section{Problem Study and \systemname Overview}
\label{sec:preliminaries}

In this section, we analyze the limitations of existing DNN partition and offloading solutions and present the problem statement with the design overview of \systemname.

\subsection{Problem Study}
Mobile applications often require \textit{fast or near real-time response} to provide better interactive experiences, examples like face recognition~\cite{chen2018mobilefacenets, martinez2021benchmarking}, health monitoring ~\cite{wu2018internet, kavitha2021iot}, and fall detection~\cite{lee2018real, yadav2022arfdnet}.
However, mobile devices (\eg smartphones, smartwatches) are always resource-constrained and cannot guarantee responsiveness for running computational-intensive algorithms (\ie DNNs). 
Recently, prior efforts on DNN compression techniques~\cite{he2018amc, deng2020model} explore reducing inference latency by trimming down model complexity, but at the cost of accuracy loss. And for large-scale DNNs, their applicability is limited.
Therefore, the \textit{DNN partition method}~\cite{kang2017neurosurgeon, li2019edge, zhang2020towards} becomes an ideal solution, which utilizes abundant computing resources of multiple edge devices to collaboratively complete the DNN computation for satisfying latency requirements without compromising accuracy.


\subsubsection{Limitations of existing DNN partition practices}
\label{sec:partitonanalysis}

The run-time deployment context is highly dynamic in three folds: \textit{(i)} the inference latency requirements may vary for diverse moments (\eg daytime and night); 
\textit{(ii)} the resource availability (\eg Cache, DRAM, and CPU/GPU compute resources) can change dynamically due to the occupancy of other applications;
\textit{(iii)} the network conditions always fluctuate dynamically.

The end-to-end DNN deployment for mobile edge computing contains two collaborative steps, \ie the \textit{DNN partition} and the \textit{partitioned module offloading} steps.
\begin{itemize}
\item  The \textit{DNN partition} step aims to find the DNN partition points with most suitable granularity to split a DNN into several modules.
The large search space for the optimal DNN partition points always results in \textit{slow search speed}. 
It thus cannot quickly adapt to the dynamically changing distributed deployment context at runtime. 
Moreover, the search space will increase with the complexity of the DNN model architecture and the number of mobile edge devices. 
\item The \textit{computation offloading} step allocates the different parts of partitioned DNN modules to diverse mobile edge devices, for satisfying latency requirements according to the dynamic resource availability of mobile edge devices. 

\end{itemize}

\begin{figure}[tbp]
    \centering
    \subfloat[Existing DNN partition practices]{
    \includegraphics[scale=0.19]{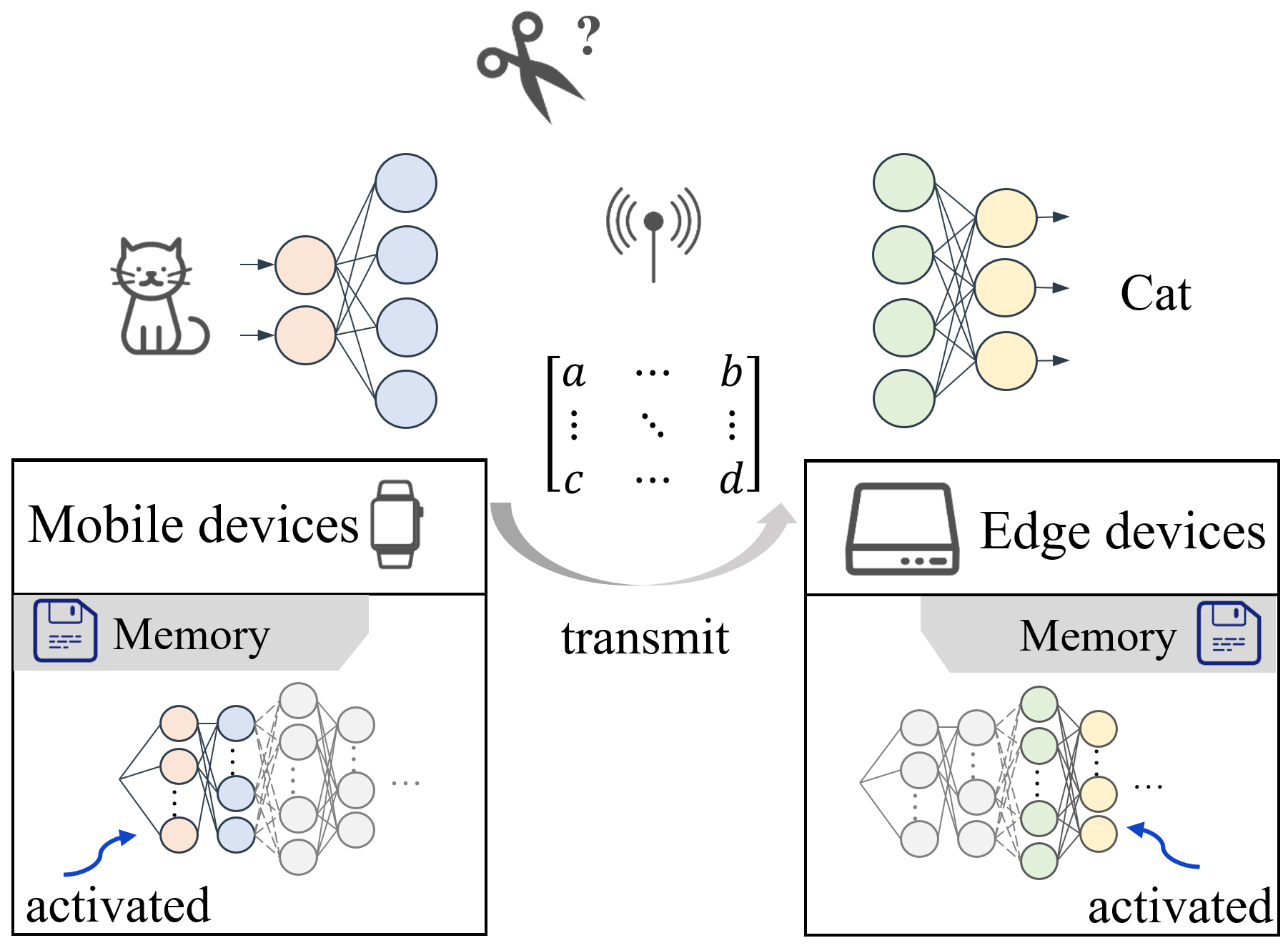}}
    \hfill
    \subfloat[Decoupled DNN partition and computation offloading]{
    \includegraphics[scale=0.145]{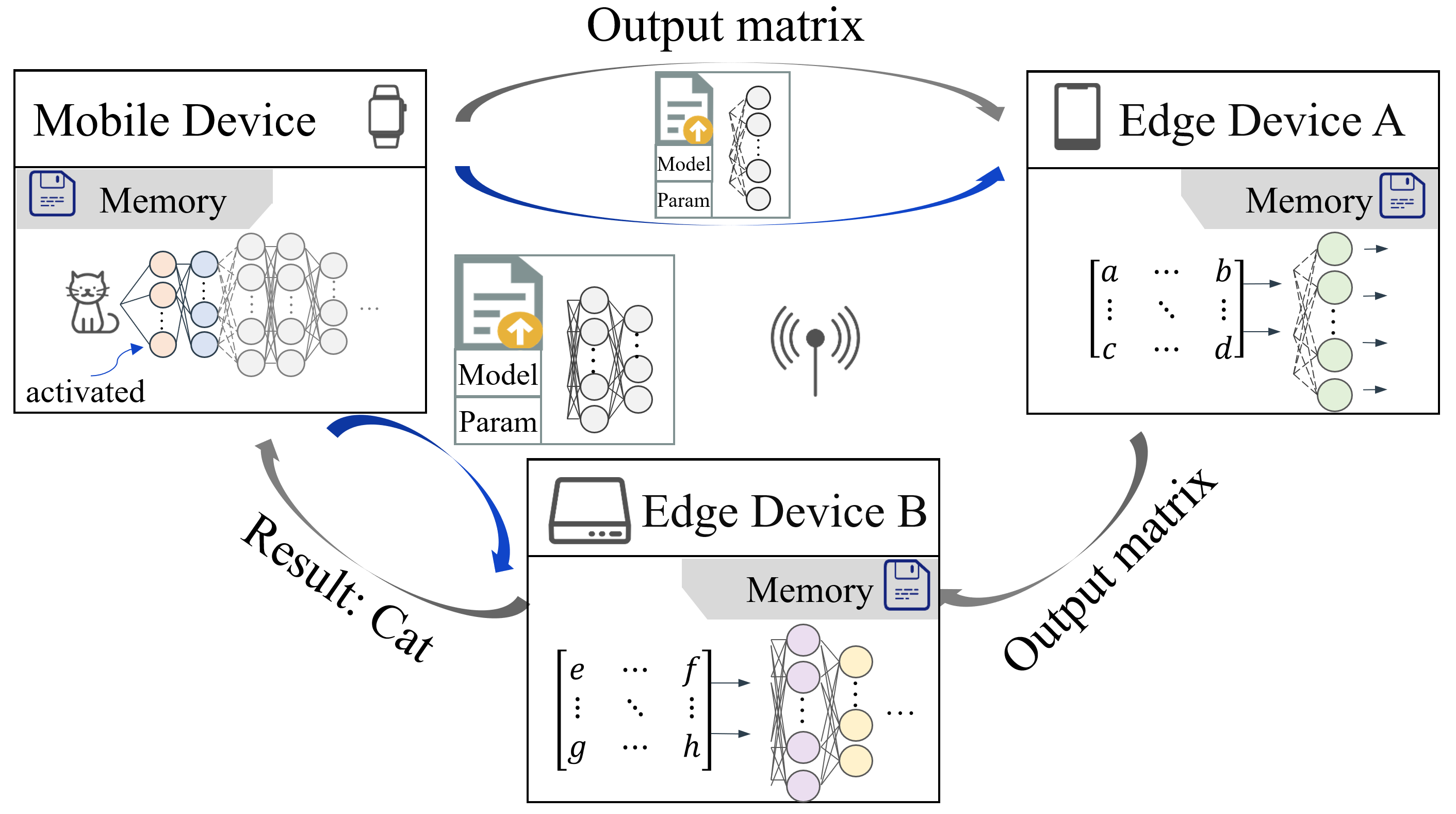}}
    \caption{Illustration of the DNN partition methods: (a) the existing coupled DNN partition practices that store full-size DNN on all devices, (b) the decoupled DNN partition and computation offloading in our work. }
    \label{img:partition_comparison}
    \vspace{-3mm}
\end{figure}

Some prior efforts couple the DNN partition step with the computation offloading step. 
For example, IONN \cite{jeong2018ionn} and SPINN \cite{laskaridis2020spinn} need to rerun both the partition and offloading steps from scratch once the deployment contexts change.
If the context changes dynamically, their significant response latency is unacceptable for latency-sensitive mobile applications.
And another type of existing research only considers how to find the best DNN partition point from the algorithmic perspective to optimize the overall inference latency, such as Neurosurgeon \cite{kang2017neurosurgeon} and DADS \cite{hu2019dads}, as shown in Fig.~\ref{img:partition_comparison}(a). 
These works are based on an assumption that all edge devices have enough memory resources to save the entire DNN parameters, and selectively compute partial modules according to the diverse partition schemes, which is not practical for resource-constrained edge devices.

\subsubsection{The Need to Decouple DNN Partition and Computation Offloading Steps}
\label{sec:moti_ofa}

As mentioned above, to efficiently adapt to the dynamic deployment context (\ie distributed resource availability, network condition, and user-specified latency requirements), we should avoid re-partition DNN or recompiling the partitioned modules every time the context changes.
This necessitates decoupling the end-to-end process of distributed DNN deployment over multiple edge devices into two independent steps, \ie DNN partition and computation offloading, as shown in Fig.\ref{img:partition_comparison}(b).
And we need the \textit{once-for-all DNN pre-partition} because:
\begin{itemize}
    \item It is independent of the application-specified latency requirements and the distributed resource availability on edge devices.
    \item It supports the prompt search for optimal computation offloading of pre-partitioned DNN modules in arbitrary and dynamic contexts.
\end{itemize}

\subsubsection{The need for runtime and accurate latency predictor}
The distributed inference \textit{latency} is an essential metric for finding the most suitable DNN deployment scheme. It includes the execution time of different partitioned modules on mobile/edge devices and the data transmission time.
Specifically, the timely and accurate latency feedback can provide a qualitative basis for determining the optimal DNN partition and computation offloading plan (detailed in $\S$ \ref{sec:predictor}).
However, it is intractable to predict the distributed inference latency because of the dynamic nature of distributed device resources and networks and the heterogeneity of DNN models will lead to ample prediction space.

\subsection{Framework Overview}
Motivated by the above analysis, we introduce \systemname, a context-adaptive and dynamically-combinable DNN deployment framework, which consists of three main functional blocks, namely \textit{once-for-all DNN pre-partition}, \textit{context-adaptive DNN atom combination and offloading}, and the \textit{runtime latency prediction}.

\textbf{Once-for-all DNN Pre-partition}.
This block aims to pre-partition the DNN model into independent modules at the fine-grained operator level, which is once-for-all and can be flexibly combined for dynamic mobile edge devices in the  context-adaptive offloading phase.
%
As mentioned in $\S$ \ref{sec:partitonanalysis}, traditional tight-coupled DNN partition algorithms need to be re-run from scratch to find the new partition scheme when the deployment context changes. 
This process leads to unacceptable responding time and degrades users' service experience. 
Previous efforts on DNN partition, such as Neurosurgeon~\cite{kang2017neurosurgeon} and QDMP~\cite{zhang2020towards}, usually leverage the DNN layer as the basic partition unit. 
It makes the large performance span between different partition schemes, causing weak adaptability to dynamic deployment context. 
There are also some advanced partition methods based on the intra-layer granular (\eg channel level~\cite{zhao2018deepthings}, operator level~\cite{teerapittayanon2017distributed}) that can provide more refined partition schemes to better adapt to the resource constraints and user requirements. 
However, the intra-layer level will lead to a massive increase in the partition scheme search space, failing to meet the real-time requirement for adaptive searching of the optimal scheme under dynamic contexts.

Therefore, \systemname decouples DNN partition and partition module offloading steps to avoid the extra overhead caused by re-partition when the deployment context changes, providing superior adaptive capabilities.
Specifically, we design a pre-partitioner to partition the DNN model at the primitive operator level to ensure flexible adaptability. 
To reduce the partition search space, we build the latency benefit function to filter partition points that do not reduce the inference latency.
And we refer to these partitioned modules as the \textit{pre-partitioned DNN atoms} in this paper, as detailed in $\S$ \ref{sec:partition}.

%
%
%
%

\textbf{Context-adaptive DNN atom combination and offloading}.
This module is designed to quickly search for the optimal combination of pre-partitioned atoms and offload them to mobile edge devices to enable fast adaptation to unknown deployment contexts (\ie resource availability, network conditions, and latency requirements).
It is non-trivial to solve such a runtime optimization problem due to multiple objectives and the vast search space of possible combinations. At the same time, the dynamic context requires the search process to be real-time and adaptive.
In \systemname, we propose the graph-based search algorithm to heuristically boost the search efficiency for the optimal combination of pre-partitioned atoms. 
Also, we present the offloading plan decision algorithm further to determine the optimal offloading order for each combination, enabling the latency benefit. 
We describe more details in $\S$ \ref{sec:offloading}.

\textbf{Runtime Latency Prediction.} 
This block provides the runtime and accurate latency feedback for the above two blocks.
Estimating the inference latency accurately at runtime is challenging because of the diversity of DNN structures and the heterogeneity of mobile edge devices.
Specifically, \systemname presents a latency predictor to estimate the overall latency for distributed computing.
We elaborately design the data sampling space for characterizing critical impact factors of DNN operators that influence the inference latency.
Furthermore, we study the relationship between latency and available memory resources. 
We observe that the dynamic device resource (\ie the available memory resource) will significantly affect the inference latency, so the dynamical fine-tuning process is proposed in the online phase to bias the predicted latency caused by the resource availability, as detailed in $\S$  \ref{sec:predictor}.
%
%

\begin{figure*}[t]
    \centering
    \includegraphics[scale=0.1]{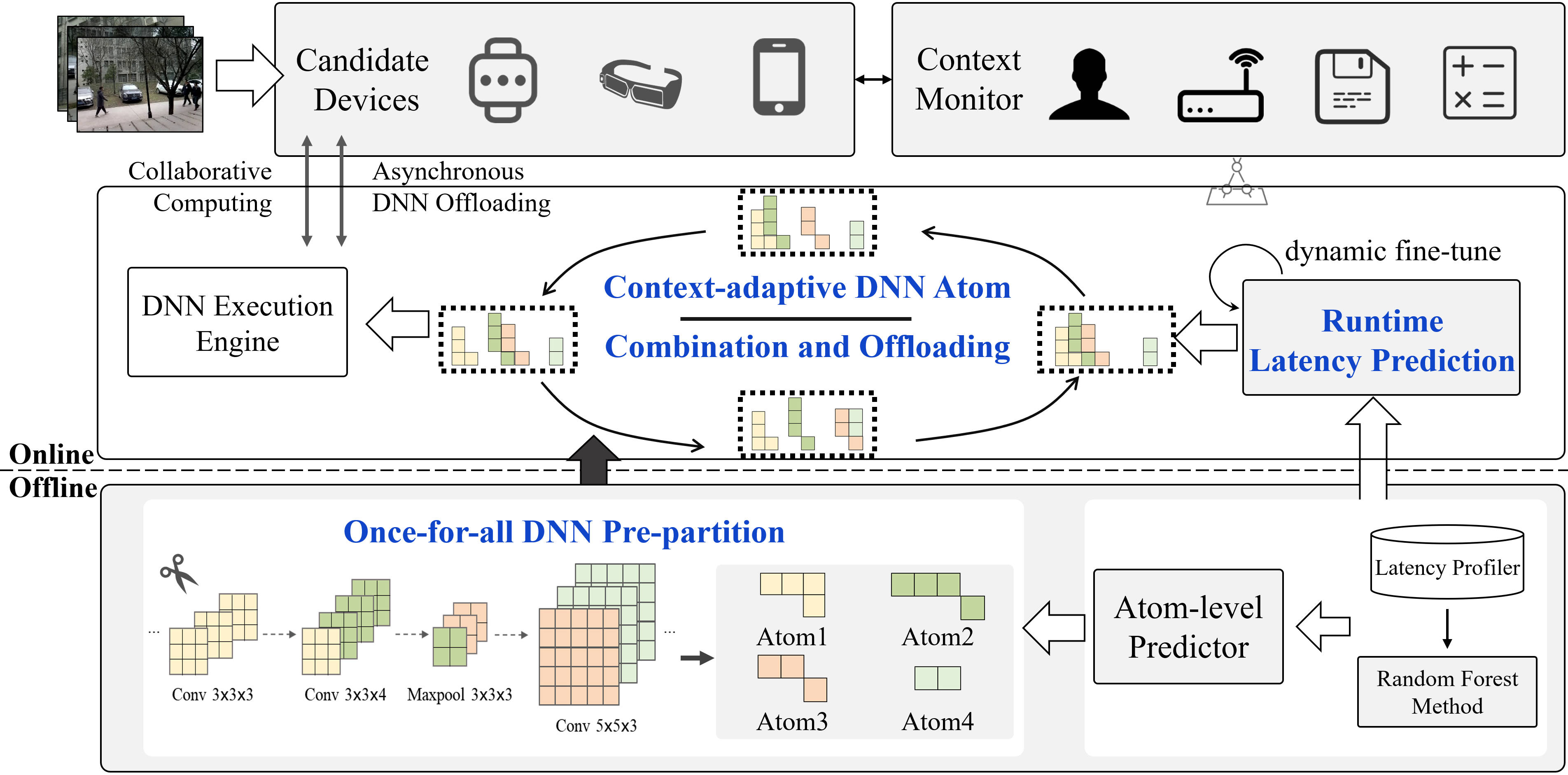}
    \caption{The system architecture of \systemname, consisting of three main blocks, once-for-all DNN pre-partition, context-adaptive DNN atom combination and offloading, and runtime latency prediction. 
    }
    \label{img:overall_architecture}
    \vspace{-3mm}
\end{figure*}

As shown in Fig.~\ref{img:overall_architecture}, the \textit{once-for-all DNN pre-partition} block generates the pre-partitioned DNN atoms.
And then, the \textit{context-adaptive DNN atom combination and offloading} block efficiently searches for the most suitable combination of pre-partitioned DNN atoms and offloads them to available edge devices. 
These atoms are in the order in which offloading benefits can be obtained the fastest for optimizing the DNN inference latency while satisfying various devices' memory and computing budgets.
And the \textit{runtime latency prediction} block provides the runtime and accurate latency feedback for the above two blocks to enable the precise module pre-partition and appropriate atom combination and offloading.

\section{DNN Pre-partition and Context-adaptive  Offloading}


In this section, we detailed describe the decoupled two phases in \systemname, \ie the \textit{once-for-all DNN pre-partition} and the \textit{context-adaptive DNN atom combination and offloading}, to achieve the context-adaptive and dynamically-combinable DNN deployment at runtime.

\begin{figure*}[ht]
    \centering
    \subfloat[Layer-level]{
    \includegraphics[scale=0.2]{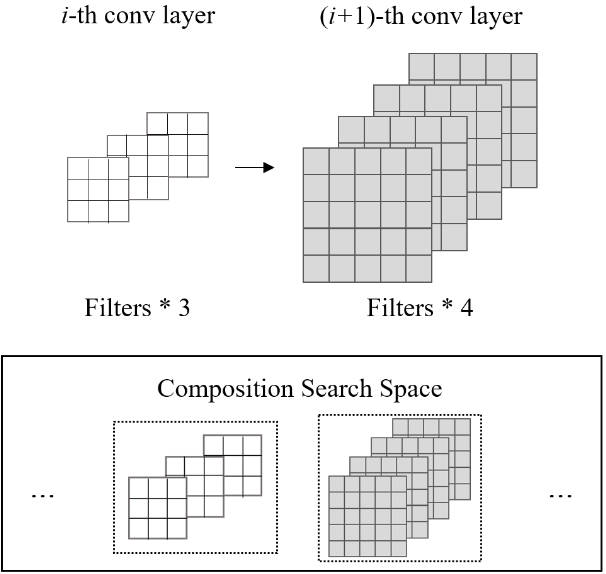}}
    \hfill
    \centering
    \subfloat[Operator-level]{
    \includegraphics[scale=0.2]{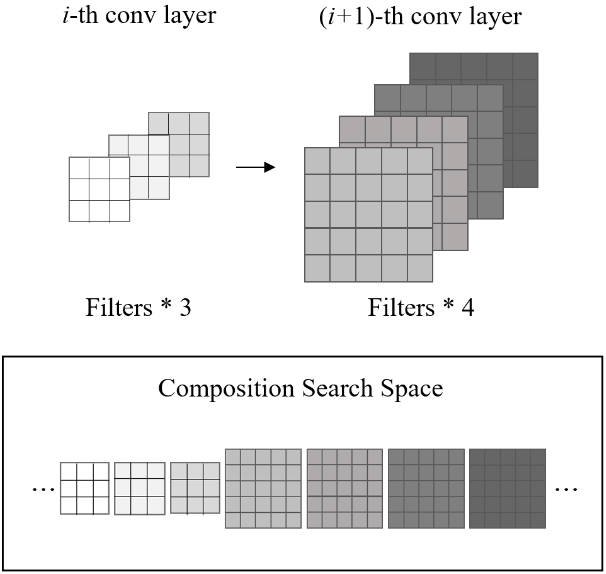}}
    \hfill
    \centering
    \subfloat[Atom-level]{
    \includegraphics[scale=0.2]{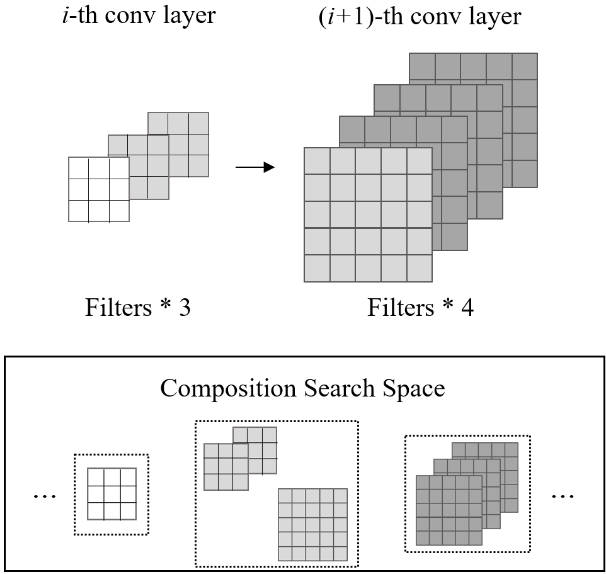}}
    \caption{Illustration of three search space granularity. 
    (a) the layer-level search space provides a limited number of possible partition schemes but lacks context adaptability, 
    (b) the operator-level search space introduces enormous search possibilities impairing search efficiency. 
    (c) our proposed atom-level search space balances the context-adaptive capabilities and the search efficiency.} 
    \label{img:prepartition}
    \vspace{-2mm}
\end{figure*} 

\subsection{Once-for-all DNN Pre-partition}
\label{sec:partition}
This section presents the design of the once-for-all DNN pre-partition block, which involves a DNN pre-partitioner to partition the mobile application-specified DNN into multiple DNN atoms based on the primitive operator and the well-designed latency benefit function.

\subsubsection{Primer on Principles of DNN Pre-partition}

%
It is non-trivial to obtain the suitable search granularity.
The coarse-grained layer-based search space in existing works, such as QDMP~\cite{zhang2020towards} and CAS~\cite{wang2021context}, makes it easy to search for the optimal partitioning scheme because of the small size of the search space. 
But the large performance span between finite partition schemes prevents them from satisfying the resource constraints and latency requirements precisely, leading to poor adaptability under the dynamic context.
The fine-grained operator-level search space, such as DeepThings~\cite{zhao2018deepthings} and DDNN~\cite{teerapittayanon2017distributed},  can provide more flexible partition schemes with higher adaptive capabilities. However, it leads to the massive expansion of the partition scheme search space.

To this end, we form a flexible and elite search space of deep model deployment by pre-partition, which includes a set of primitive atom-level DNN modules.
Based on the pre-partitioned DNN modules, we can flexibly combine them into different computation offloading schemes at runtime to deal with dynamically changing deployment contexts.
Our key idea for atom-level DNN pre-partition is mainly based on the following two principles:

\begin{itemize}
    \item  \textit{Pre-partitioning the DNN based on primitive operators for Narrow and Flexible Candidate Searching Space}. 
    Although the heterogeneity of DNNs will lead to a large search space for the optimal model partition scheme, the primitive operators (\eg convolution, pooling, and fully connection operators) of the DNN are in a stable and narrow space. 
    Therefore, we first form the search space based on the primitive operator level to improve the searching convergence speed.
    %
    %
    \item \textit{Filtering Pre-partition Candidate Points for Elite Searching Space}.
    We then filter the above-obtained candidate partition points to reduce the search space further.
    We find that not all primitive operators of the model are suitable to serve as the DNN partition points. 
    As we will show in $\S$ \ref{sec:experiment}, we experimentally observe that some operators are almost impossible to be selected as the partition points due to the large volumes of output feature maps and huge transmission cost.
    These findings are also indicated in \cite{jeong2018ionn, wang2021context}.
    Thus, to reduce the search space and boost the subsequent search efficiency, we further filter the candidate partition points based on whether partitioning at this point can reduce the total inference latency and satisfy the constraints.
\end{itemize}

Fig.\ref{img:prepartition} shows a deployment example of AlexNet (including 5 convolutional layers and 3 fully connected layers) on three available mobile edge devices.
AlexNet is partitioned into 8 modules in the layer-level search space, which can be combined into $3^{8}$ deployment schemes.
However, the layer-based offloading schemes are less adaptive to dynamic contexts.
%
The operator-level search space can generate $3^{23}$ deployment schemes, which is not conducive to the real-time performance of optimization.
At the atom level, there are $3^5$ possible deployment schemes.
And it can balance context-adaptive adaptability and search efficiency.

\subsubsection{Latency Benefit-guided Pre-partition}
To precisely measure the latency benefit of the above-obtained candidate DNN partition points and the subsequent 
combinational offloading schemes, we define a \textit{latency benefit function}.
This function considers the reduction of overall inference latency and the satisfaction of the user-defined latency demands. 
The latency benefit function is defined as follows.

\begin{align}
R_{off}(p_{i}) = \lambda_{1}log\frac{T_{exe}(p_{i})-T_{dev}}{T_{tran}(p_i)} -\lambda_{2}[\varepsilon(T_{exe}(p_i) + T_{tran}(p_i)-T_{user})]
\label{benefit_off}
\end{align}
where $p_{i}\in P$, $P$ represents the set of DNN partition schemes determined by all possible partition points.

The latency benefits $R_{off}$ are composed of the benefit of inference latency reduction and the benefit of satisfying user requirements, weighted by $\lambda _{1}$ and $\lambda _{2}$.
In particular, the inference latency benefits $R_{off}$ consists of two parts: the acceleration benefit $T_{exe}(p_{i})-T_{dev}$ brought by collaborative devices and the negative latency benefit $T_{tran}(p_i)$ caused by data transmission.
$T_{exe}$ indicates the distributed execution latency of $p_{i}$ on both edge and mobile devices, and $T_{dev}$ represents the latency of completely executing the model on the local side of the mobile device. 
Thus the difference between them represents the latency acceleration benefit of offloading a DNN on a specific point.
The execution latency of any DNN module can be predicted by our designed runtime latency predictor, detailed in $\S$ \ref{sec:predictor}.It is worth mentioning that the objects manipulated during time latency prediction in pre-partition are fine-grained than during atom combination and offloading, because this is the first time the DNN is partitioned.
And we adopt logarithmic operations (\ie $log$ in Formula ~\eqref{benefit_off}) to reduce the computational complexity.

Moreover, we denote the user requirements on the overall latency budget as $T_{user}$.
When the latency $T_{exe}(p_i)+T_{tran}$ is less than $T_{user}$, we define $\varepsilon (T_{exe}(p_i)+T_{tran}(p_i)-T_{user}))=0$, where $\varepsilon$ is a unit-step function.
Conversely, if the latency budget is exceeded, the partitioned modules will get a negative benefit.

In detail, the distributed execution latency $T_{exe}$ contains the computation latency at the local side of the mobile device and the $N$ edge devices, calculated as follows:

\begin{eqnarray}
T_{exe}(p_i) = t_{local} + \sum_{j}^{N} t_{edge}^{j}
\label{equ:latency}
\end{eqnarray}
where $t_{edge}^{i}$ is the execution latency of the combined module offloaded to the $j-th$ edge device.

And the data transmission latency $T_{tran}(p_i)$ consists of the time consumption of transmitting the intermediate data among mobile device and edge devices, as shown follows:

\begin{eqnarray}
T_{tran}(p_i) = (D_{local}+\sum_{j}^{N} D_{edge}^{j})/B(t) 
\label{equ:tran_latency}
\end{eqnarray}
where $D_{local}$ and $D_{edge}^{j}$ represent the data size of intermediate feature maps that the local device and the $j-th$ edge device need to transmit, respectively. And $B(t)$ is the time-variant network bandwidth.

%

Upon the above criterion, we compute the latency benefits of all candidate partition points of the task-specific DNN and filter out the points with negative latency benefits. 
And then, the DNN is partitioned into multiple modules in the positions using the filtered partition points. 
In this way, the obtained partition modules, which are defined as \textit{pre-partitioned DNN atoms}, are in a minimized and valid search space, as shown in Fig.\ref{img:prepartition}(c).
And the offloading of any atom combinations can bring positive latency benefits by filtering inappropriate partition points.
And then, we store these pre-partitioned atoms as executable model files based on the deep learning framework (\ie the pkl file format in Pytorch) so that the mobile and edge devices can load and execute them uniformly.
Specifically, we define subclasses of $torch.nn.module$ for all pre-partition atoms to regularize them into executable neural network classes in Pytorch and modify the $forward$ functions to implement the forward propagation of data for calculating the output features of atoms. 
In addition to the DNNs with chain structure (\eg ResNet, VGG), our pre-partition mechanism can be further applied to other advanced DNNs with DAG structure (\eg GoogLeNet), since they are also the combination of primitive operators.

\subsection{Context-adaptive DNN Atom Combination and Offloading}
\label{sec:offloading}
To adaptively offload the appropriate atom combinations on corresponding devices for satisfying the dynamic contexts, we establish a search graph for  DNN atom combinations and propose a fast decision algorithm to achieve the most suitable DNN atom combination and offloading strategies efficiently.

\subsubsection{Optimization Problem Formulation}
Choosing the optimal combination for adapting to the dynamic mobile edge contexts is intractable, considering a complex DNN model may contain massive combinations of pre-partitioned atoms with varying resource budgets.
We formulate the DNN atom combination and offloading problem as a runtime optimization problem.
Specifically, \systemname expects to search for the optimal combination $s_{opt}$ based on pre-partitioned DNN atoms to maximize the latency benefits $R_{off}$ under the constraints of the user-specified latency requirements and the resource (\ie memory and computing) budgets of mobile and edge devices.
Formula ~\ref{equ:optimizeproblem} shows the problem definition.

\begin{eqnarray}
\label{equ_optimizeproblem}
& \mathop{\arg\max} \limits_{s_i \in S} R_{off}(s_{i}) 
 & \begin{array}{r@{\quad}r@{}l@{\quad}l}
s.t.&
\left\{\begin{matrix} T_{exe}(s_i)+T_{tran}(s_i) \le  T_{user}(t)\\ 
C^{j}(s_i) \le C_{budg}^{j}(t) & j\in [0,n] \\
M^{j}(s_i) \le  M_{budg}^{j}(t) & j\in [0,n] \end{matrix}\right.
\end{array}
\label{equ:optimizeproblem}
\end{eqnarray}
where $s_i \in S$, $S$ is the set of all possible combinations of pre-partitioned atoms. $T_{exe}(s_i)+T_{tran}(s_i)$ denotes the total inference latency of $s_i$, which should meet the user latency requirement of $T_{user}(t)$. We express the dynamic deployment contexts as time-varying constraints, \ie the memory budget $M_{budg}(t)$ and the computation budget $C_{budg}(t)$. $C^{j}(s_i)$ represents the computation amount for running the target atoms on the device $j$, which is measured by the floating-point operations (\ie FLOPs). Similarly, $M^{j}(s_i)$ is the required memory space for pre-partitioned atoms on device $j$.


\subsubsection{Construction of Search Graph}
To formally represent the search space of combinations, we establish a search graph that contains multiple configurations and performance indicators when deploying atoms on multiple devices.
Specifically, given a DNN with pre-partitioned atoms, we can obtain all possible combination offloading schemes and organize them as the graph $G=<V, L>$. 
Each vertex $v_{i}$ in $G$ represents an optional combination offloading scheme $s_i$, marked with its inference latency $T(s_i)$, the distributed memory demands $M(s_i)$, and computation amount $C(s_i)$.
And a connect $<v_{i}, v_{j}>$ $\in$ $L$ between $v_{i}$ and $v_{j}$ indicates that only one atom in the combinations of $v_i$ and $v_j$ is in a distinct order.
For example, $v_i$ is obtained by offloading atom 2 to the edge, and $v_j$ is obtained by offloading atom 2 and atom 3.

\begin{figure}[tbp]
    \centering
    \includegraphics[scale=0.35]{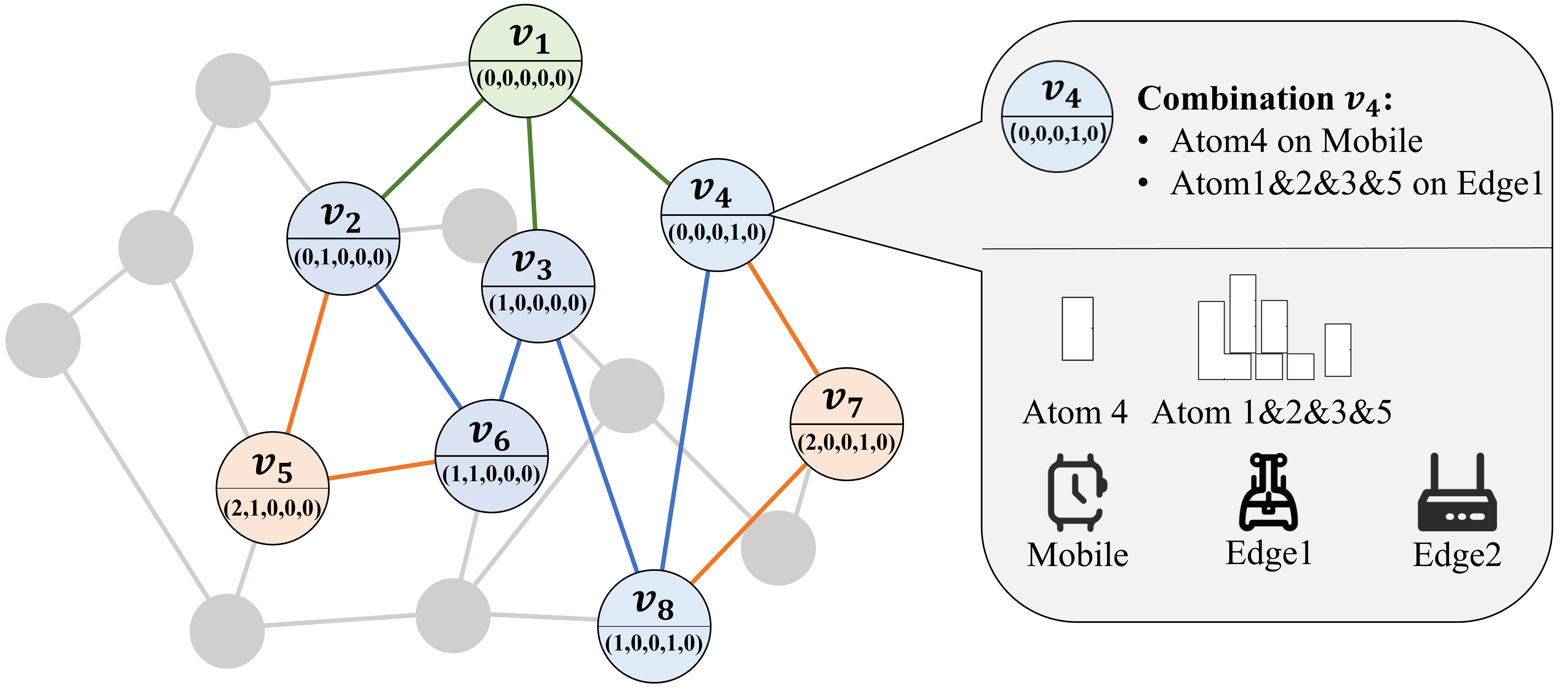}
    \caption{Illustration of search graph $G$ for deploying AlexNet with five atoms on the local side of mobile device and two edge devices. The different colors indicate the different number of deployed devices in the combination offloading scheme.}
    \label{img:patitionstategrpah}
    \vspace{-2mm}
\end{figure}

As shown in Fig.~\ref{img:patitionstategrpah}, we present an example of how to construct a search graph based on the atoms of AlexNet (a typical DNN consisting of five atoms) to aid understanding.
It illustrates the constructed search graph deployed on two edge devices (\ie $edge1$ and $edge2$).
The vertex also marks with a vector parameter, which reflects how the atoms are combined and offloaded to edge devices. For example, the vertex $v_{3}$ with vector parameter $(1,0,0,0,0)$ indicates that $atom2$, $atom3$, $atom4$, and $atom5$ are offloaded to $edge1$, and $atom1$ is executed on the local side of the mobile device. Once the deployment context changes resulting in a mismatch between the scheme resource requirements and performance budgets, \systemname will search for a new combination offloading scheme in the graph (as modeled in Equation~\ref{equ:optimizeproblem}).


\subsubsection{Context-adaptive Combination of Pre-partitioned DNN Atoms}
Our goal with the search graph is to minimize the search time for fast and efficient adaptation to the deployment context. 
Therefore, we propose the context-adaptive decision algorithm to search for the optimal combination of pre-partitioned atoms iteratively in a progressively increasing subspace rather than the complete space \cite{kang2017neurosurgeon, li2019edge}.

We map the search graph into a three-dimensional solution space composed of computation amount, memory footprint, and inference latency. 
And the combination offloading problem can be divided into two sub-problems: (1) finding the combination offloading subspace that satisfies the strict constraints and (2) searching the optimal combination of pre-partitioned atoms with maximum offloading benefit. 
Inspired by the prior work \cite{wang2021context}, two adjacent combination solutions have similar performance.
If a combination satisfies the constraints, its adjacent combinations will also meet them with a high probability. 
These adjacent points form a search subspace.
Accordingly, we propose an iterative search algorithm and construct the "artificial gradient'' to heuristically guide the search process for fast convergence.

\begin{figure}[tbp]
    \centering
    \includegraphics[scale=0.42]{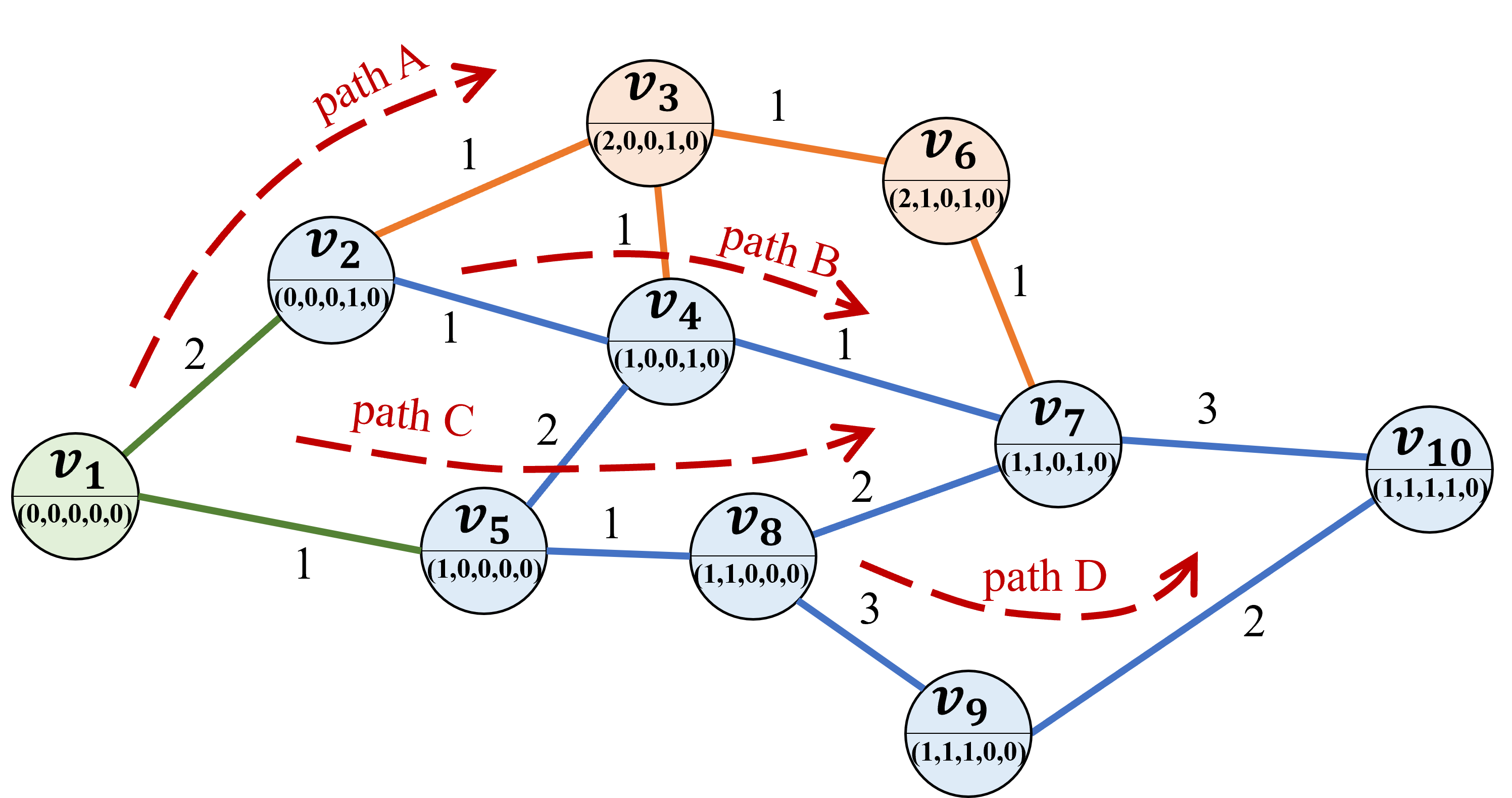}
    \caption{Illustration of four possible combination offloading paths from $v_1$ to $v_{12}$. $v_1$ is the current combination of atoms, and $v_{12}$ is the target combination searched by our designed algorithm.}
    \label{img:offloading_paths}
    \vspace{-2mm}
\end{figure}

In detail, the euclidean gap $d$ measures the distance (\ie loss function) between a specific combination of pre-partitioned atoms and the constraints of the new context (or another variety) , as shown in Formula ~\ref{equ:distance}.

\begin{align}
\nonumber
d(s_i, C_t)= [\alpha (T_{exe}(s_i)+T_{tran}(s_i)-T_{user}(t))^{2} + 
\\ \beta \sum_{j}^{k}(C^j(s_i)-C_{budg}^j)^{2} + \gamma \sum_{j}^{k}(M^j(s_i)-M_{budg}^j)^{2}]^{\frac{1}{2}}
\label{equ:distance}
\end{align}
where $\alpha$, $\beta$, and $\gamma$ represent the priorities to satisfy the constraints $C_t$, \ie the latency requirement, the computation amount, and the memory footprint constraints.

When the deployment context or latency requirements change, the current combination of pre-partitioned atoms may no longer satisfy hard constraints. 
Thus our algorithm is triggered to find a new combination on the fly. 
The main process of the \textit{context-adaptive combination decision algorithm} is illustrated as follows:
(1) We initialize $Set$ with the current combination $v_{cur}$. (2) The adjacent combination of $v_i$ in $Set$ will be expanded into $Set$. (3) We filter and update the top $k$ closest combinations to the target constraints. The choice to keep $k$ instead of just the closest combination is to avoid getting stuck in a local optimum. (4) If $v_i$ does not satisfy the target constraints, change the next combination and return to (2). 
If any combination in $Set$ satisfies the target constraints, the iteration's goal switches to find the variety with maximum benefit. (5) $Set$ is repeatedly expanded to find the maximum offloading benefit until it remains constant and the constraints of the target context are satisfied.

\subsubsection{DNN Atom Combination Offloading}

To ensure that the offloading benefits can be obtained as soon as possible and avoid additional offloading overheads, we decide on the combination offloading plan based on the shortest path algorithm to determine the optimal offloading order for each atom in the target combination.

The offloading order is notably important within the specific combination of pre-partitioned atoms. 
If we set the order randomly, the atom that brings a slight latency benefit may be offloaded first and defeat our purpose.
For example, as shown in Fig.~\ref{img:offloading_paths}, a path from the current combination $v_1$ to the target $v_{10}$ can be considered as an offloading plan, \eg $pathC$ denotes the offloading plan of atoms as $atom1$, $atom2$, $atom3$, and $atom4$. 
The weight on the connection represents the transmission latency required to offload the atom to the target device.
$pathA$ offloads atoms to $edge2$, which will not actually participate in the final collaborative computing determined by $v_{10}$, resulting in extra transmission overhead.
And for pathB, the mobile device requires $2s$ to obtain the offloading benefit of the first atom, which needs a longer waiting time than pathC.

To this end, we design the offloading plan decision algorithm based on the shortest path strategy to find the optimal offloading order for each DNN atom combination.
Specifically, the design of our algorithm follow the following two principles. 
First, we prioritize offloading the atom with the lowest overhead (\ie the transmitting latency caused by offloading network structures and parameters to edge devices) at each decision step, to engage edge devices early in the collaborative computing process.
Second, the overall offloading overhead from $v_{cur}$ to $v_{tar}$ should be the lowest to ensure that there is no unnecessary offloading.

\begin{algorithm}[t]
\begin{flushleft}
\textbf{Input:} the current combination offloading scheme \textbf{$v_{cur}$}, the target scheme \textbf{$v_{tar}$}, and the graph \textbf{$G$}.
\end{flushleft}
\begin{flushleft}
\textbf{Output:} the offloading plan $p_{off}$ for required modules
\end{flushleft}
\caption{Offloading Plan Decision}
\label{alg:offloading}
\begin{algorithmic}[1]
\State \textbf{function} OffloadingPlanDecision($v_{cur}$, $v_{tar}$)
\State Initialize auxiliary
array $L$, and the overhead of each scheme $v_i$
\For{$v_i$ $\in$ $G$}
\If{there is no $L$ between $v_{cur}$ and $v_{i}$}
\State Set $L[i]$ to $INF$
\Else 
\State Set $L[i]$ to the overhead of $v_{i}$
\EndIf
\EndFor
\State Add $v_{cur}$ into $U$
\State Expand the adjacent combinations of $v_{cur}$ into $Q$
\While{$U$ != $V$}
\State Pick $v_{temp}$ with the minor overhead in $Q$
\For{$v_j$ in the adjacent combinations of $v_{temp}$}
\State Calculate the new overhead updated by $v_{temp}$
\If{$overhead_{new}$ $<$ $L_{j}$}
\State ${L_{j}}$ $\longleftarrow$ $overhead_{new}$
\EndIf
\EndFor
\State Add $v_{temp}$ into $U$, and exclude it from $Q$
\State Expand the adjacent combinations of $v_{temp}$ into $Q$
\EndWhile
\State $p_{off}$ $\longleftarrow$ the migration path from $v_{cur}$ to $v_{tar}$ with the least overhead 
\State return $p_{off}$
\end{algorithmic}
\end{algorithm}

Algorithm\ref{alg:offloading} illustrates the offloading plan decision algorithm. 
We first initialize the search graph connections with different weights to represent the offloading overheads (\ie the transmitting latency) of the DNN atoms between two adjacent combinations. 
The overhead from $v_{i}$ to $v_{j}$ is the sum of overheads on the offloading path. 
And then, the offloading plan decision algorithm includes the following steps: (i) An auxiliary array $L$ is initialized, in which each element $L[i]$ represents the currently updated minimal overhead from $v_{cur}$ to vertex $v_{i}$. 
If a connection exists from $v_{cur}$ to $v_i$, $L[i]$ is the weight, otherwise $L[i]=\infty$ (line2 - 9).
(ii) The $v_{cur}$ is added to the set $U$, which denotes the set of vertices with the currently determined lowest overhead to $v_{cur}$. 
We add the vertices adjacent to $v_{cur}$ to the set $Q$, which denotes the set of vertices with the undetermined lowest overhead to $v_{cur}$ (line10, 11). 
(iii) The vertex $v_{temp}$ with the minor overhead to $v_{cur}$ in $Q$ is selected to traverse its adjacent vertices. 
We update the values of reached vertices in $L$ if the newly obtained overheads are lower than the original overheads (line13 $\sim$ 19). 
(iv) We add $v_{temp}$ to $U$ and the newly updated vertices to $Q$ (line20 $\sim$ 21);
(v) Repeating (3) and (4) until $U$ contains all vertices. Afterward, the path with the 
least overhead from $v_{cur}$ to $v_{tar}$ is our expected optimal offloading plan (line22 $\sim$ 24). 
Once the offloading plan is determined, \systemname will offload executable model files (\ie .pkl) to target devices in order. 
\section{Runtime Latency Prediction}
\label{sec:predictor}

The DNN pre-partition and the dynamic combination and offloading of atoms are all sensitive to the inference latency of deep model on the physical device. 
Therefore, in this section, we present the runtime latency predictor to accurately and timely predict the inference latency (including the execution latency and transmission latency) for diverse DNN partition and combination offloading schemes at runtime. 
We can directly obtain the transmission latency by dividing the transmitted data by the current network bandwidth. Nevertheless, the execution latency is more challenging to predict due to the variability of the partitioned model and the heterogeneity of the computing devices.
The focus of our predictor is to model the complex correlations between the DNN configuration parameters (\eg kernel size, input channel number) and the corresponding execution latency (\ie the latency for distributed execution), and also to capture the latency bias caused by the dynamic changes of edge resource availability (\eg available memory resource).

\subsection{Data Sampling for Latency Prediction}
\begin{figure}[t]
  \centering
  \subfloat[Height/width of conv]{
  \includegraphics[height=0.178\textwidth]{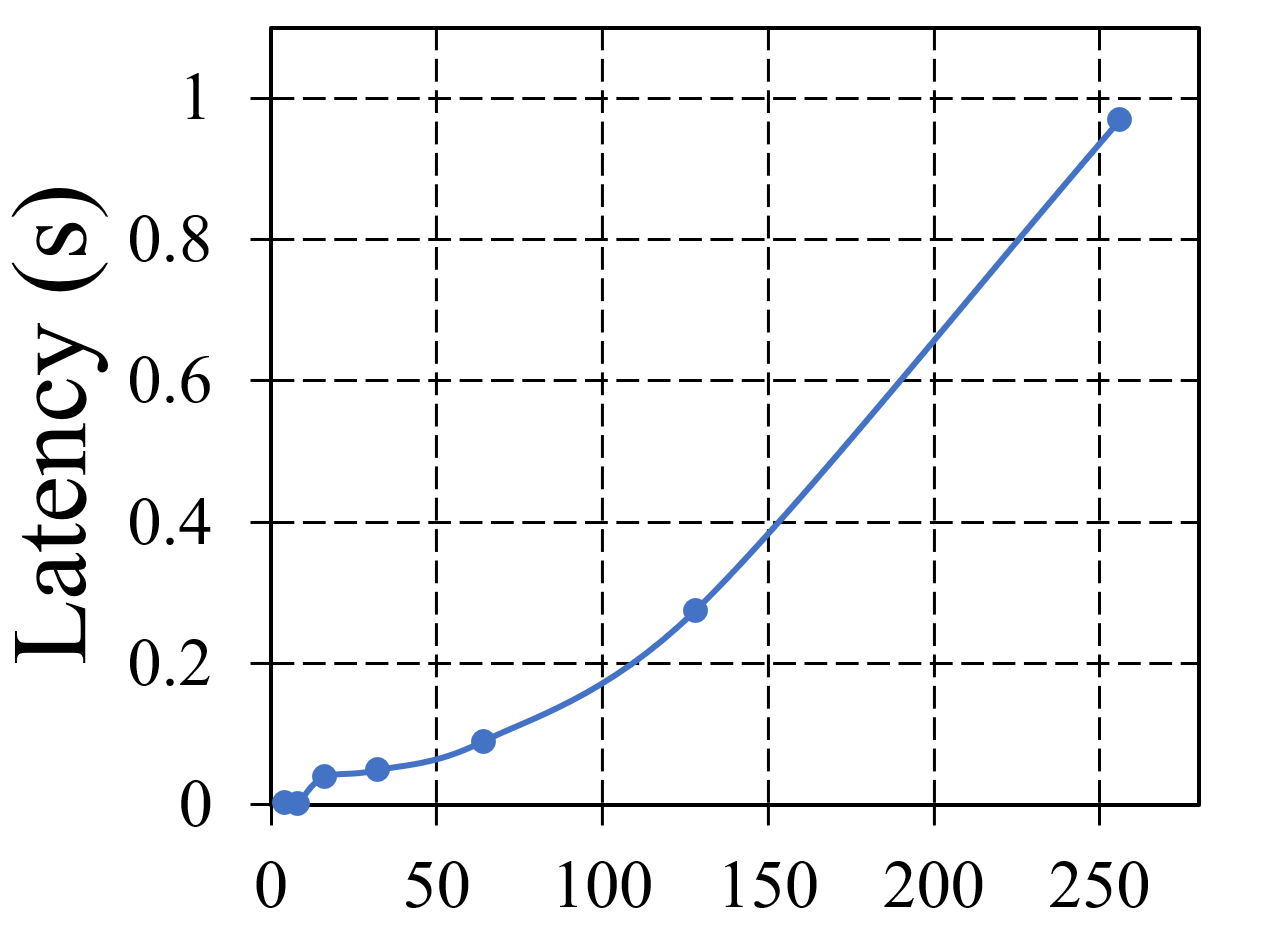}}
  \hfill
  \subfloat[Size of conv kernel]{
  \includegraphics[height=0.178\textwidth]{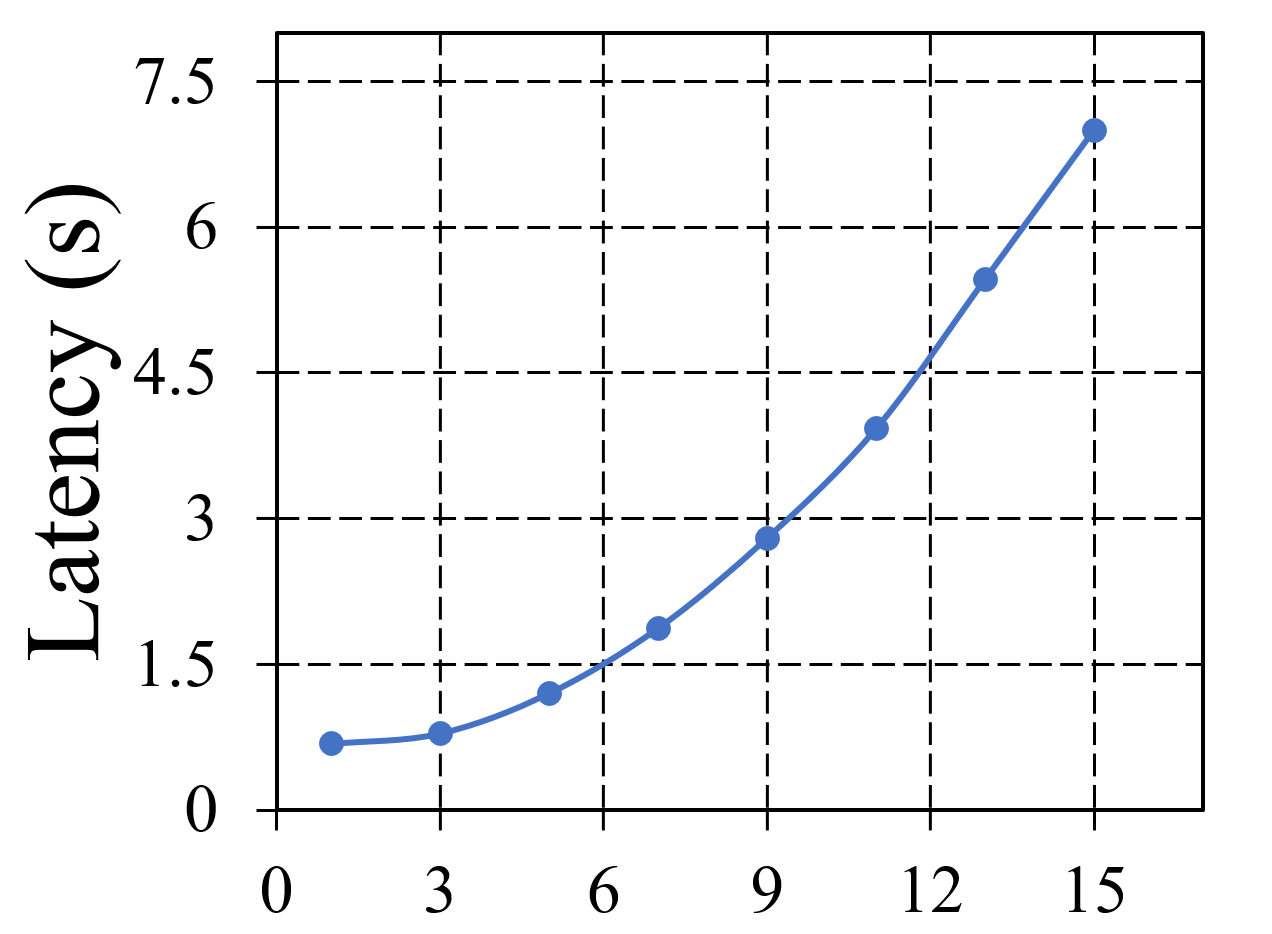}}
  \hfill
  \subfloat[In channels of conv]{
  \includegraphics[height=0.178\textwidth]{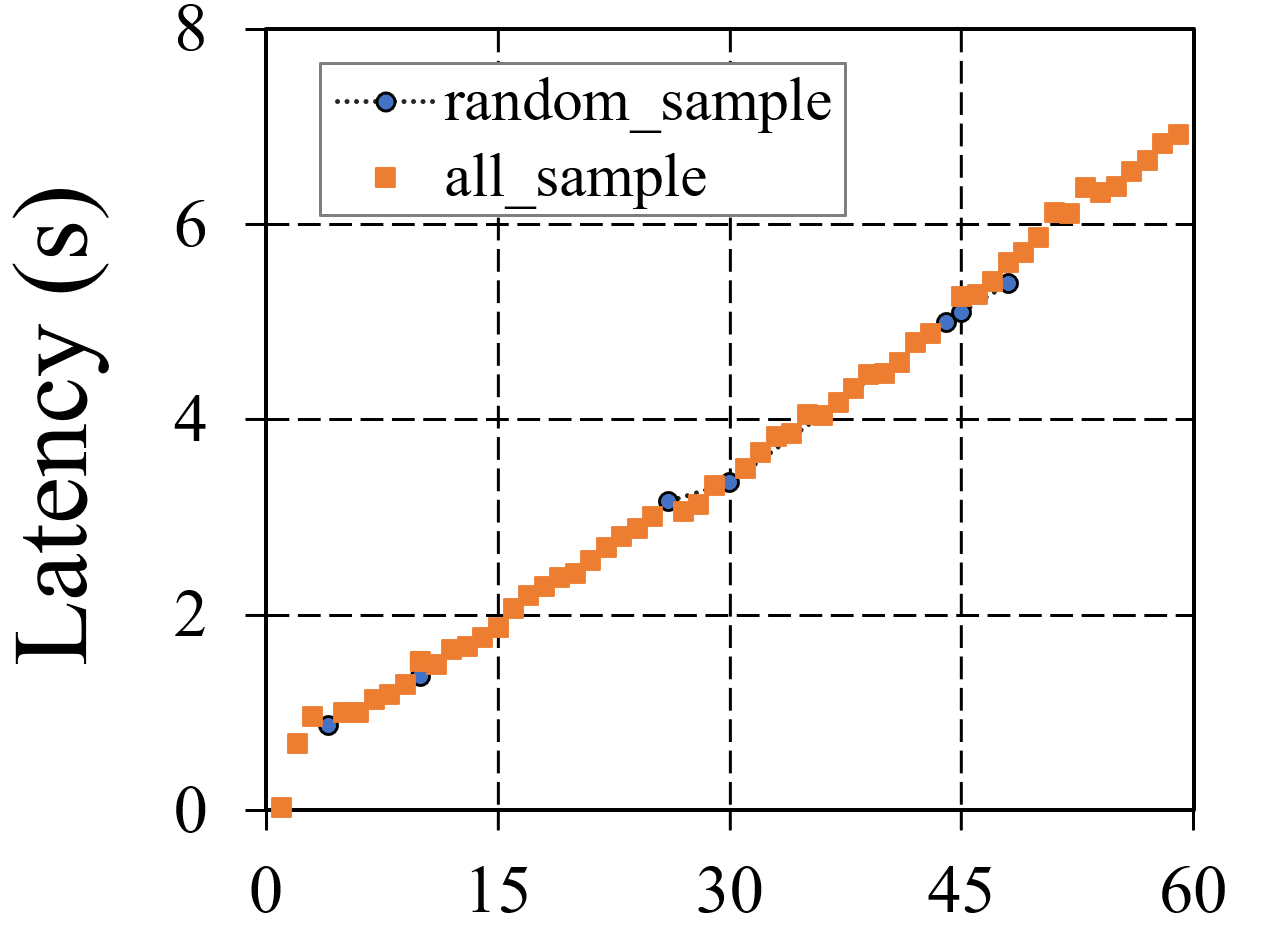}}
  \hfill
  \subfloat[Out channels of conv]{
  \includegraphics[height=0.178\textwidth]{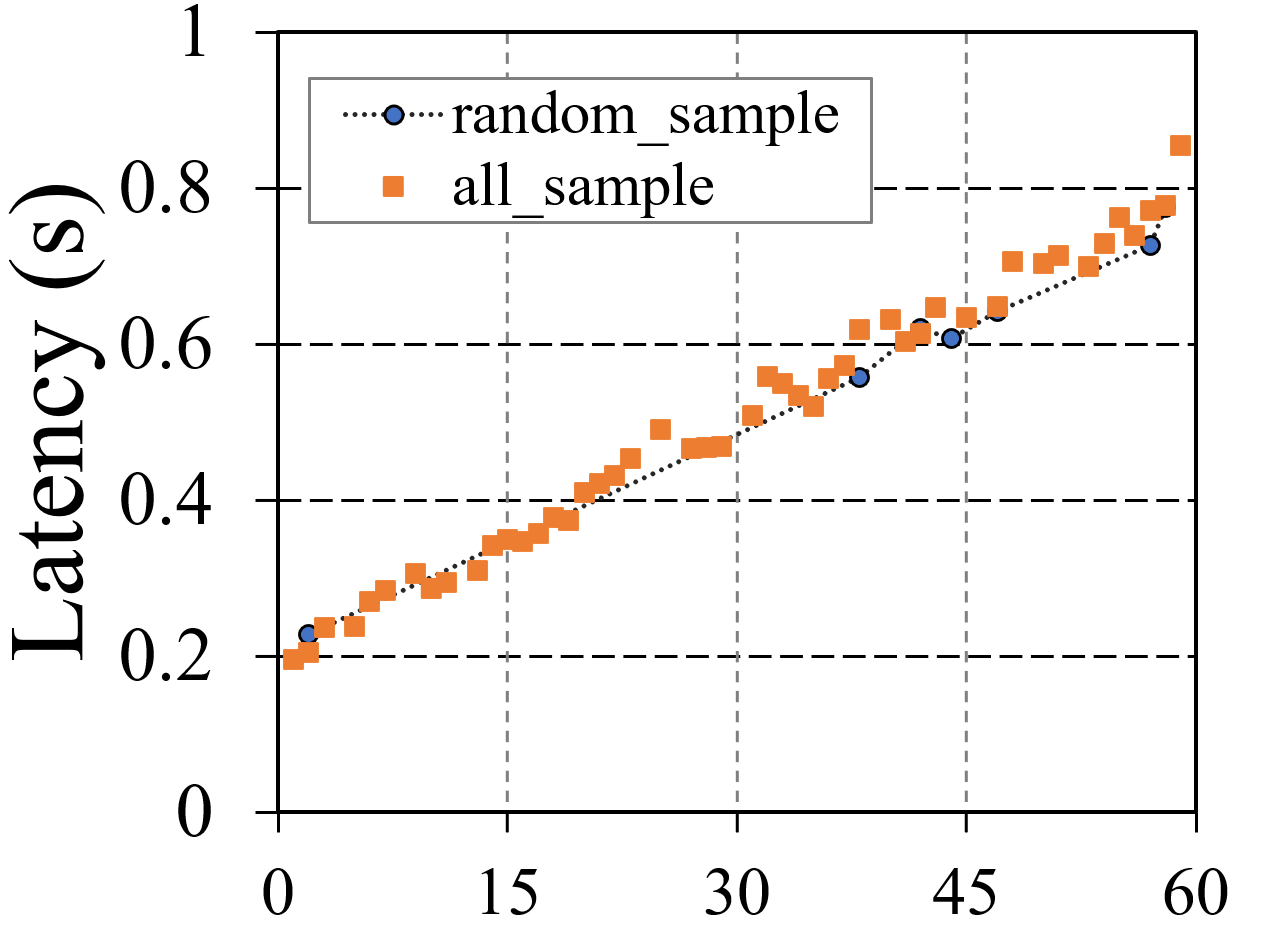}}
\vspace{-1mm}
\caption{The inference latency changes of the typical model operator (Conv) under different configuration parameters.}
\label{img:latency_patterns}
\vspace{-3mm}
\end{figure}

We first demonstrate the challenges of predicting the execution latency of DNNs with complex and variable structures and introduce the design of data sampling space.

\subsubsection{The Non-linearity Relationship between the Execution Latency and Model Configuration Parameters} 
\label{sec_nonlinearity}

It is non-trivial to predict the execution latency for \textit{various} DNN structures with diverse computing devices.
Previous works~\cite{kang2017neurosurgeon, li2019edge} usually predict the DNN execution latency at a single device, based on the assumption that there is a linear relationship between the DNN complexity and the latency. However, we experimentally find that the execution latency shows a non-linear relationship with the DNN complexity, complicating the latency prediction.

Specifically, to mine the latency patterns, we sample various configuration hyperparameters (\eg the height and width of the feature map, kernel size, input and output channels) of typical operators (\eg convolutional (Conv), and fully-connection (FC)), and measure the corresponding execution latency on the Raspberry Pi 4B as an example. 
As shown in Fig.~\ref{img:latency_patterns}, different operator configurations affect the execution latency in distinct patterns. 
For example, the latency of Conv is proportional to the square of the height and width of the input feature map, indicated in Fig.~\ref{img:latency_patterns}(a)(b). While the size of input and output channels exhibits the staircase correlation with the latency, as shown in Fig.~\ref{img:latency_patterns}(c)(d)). 
%
Therefore, we should carefully design the sampling space and sampling method to train the latency predictor with high accuracy and generalization ability with limited data.

\subsubsection{The Design of Data Sampling Space}
\begin{figure*}[t]
  \centering
  \subfloat[AlexNet]{
  \includegraphics[height=0.236\textwidth]{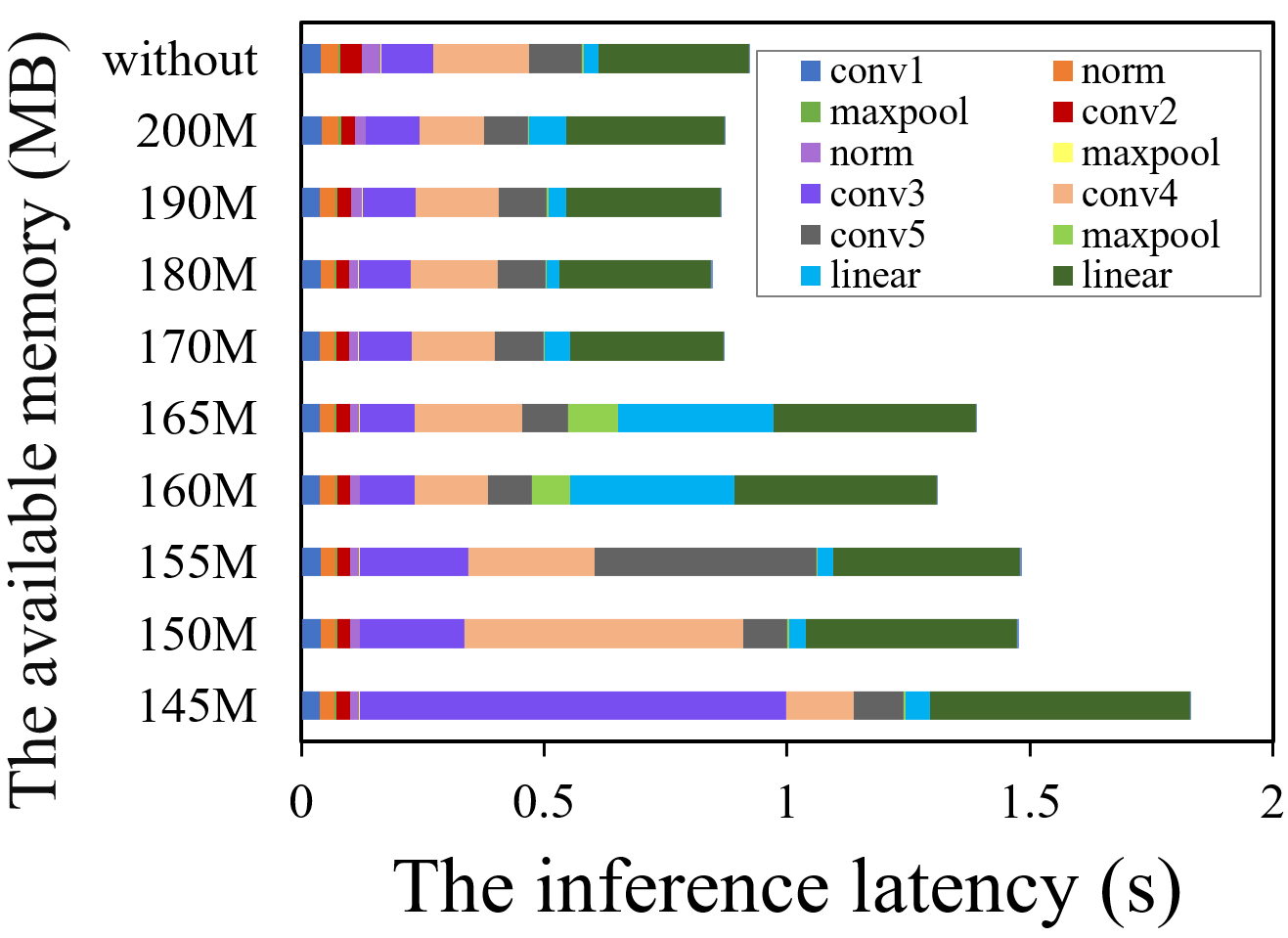}}
  \hfill
  \centering
  \subfloat[VGG]{
  \includegraphics[height=0.236\textwidth]{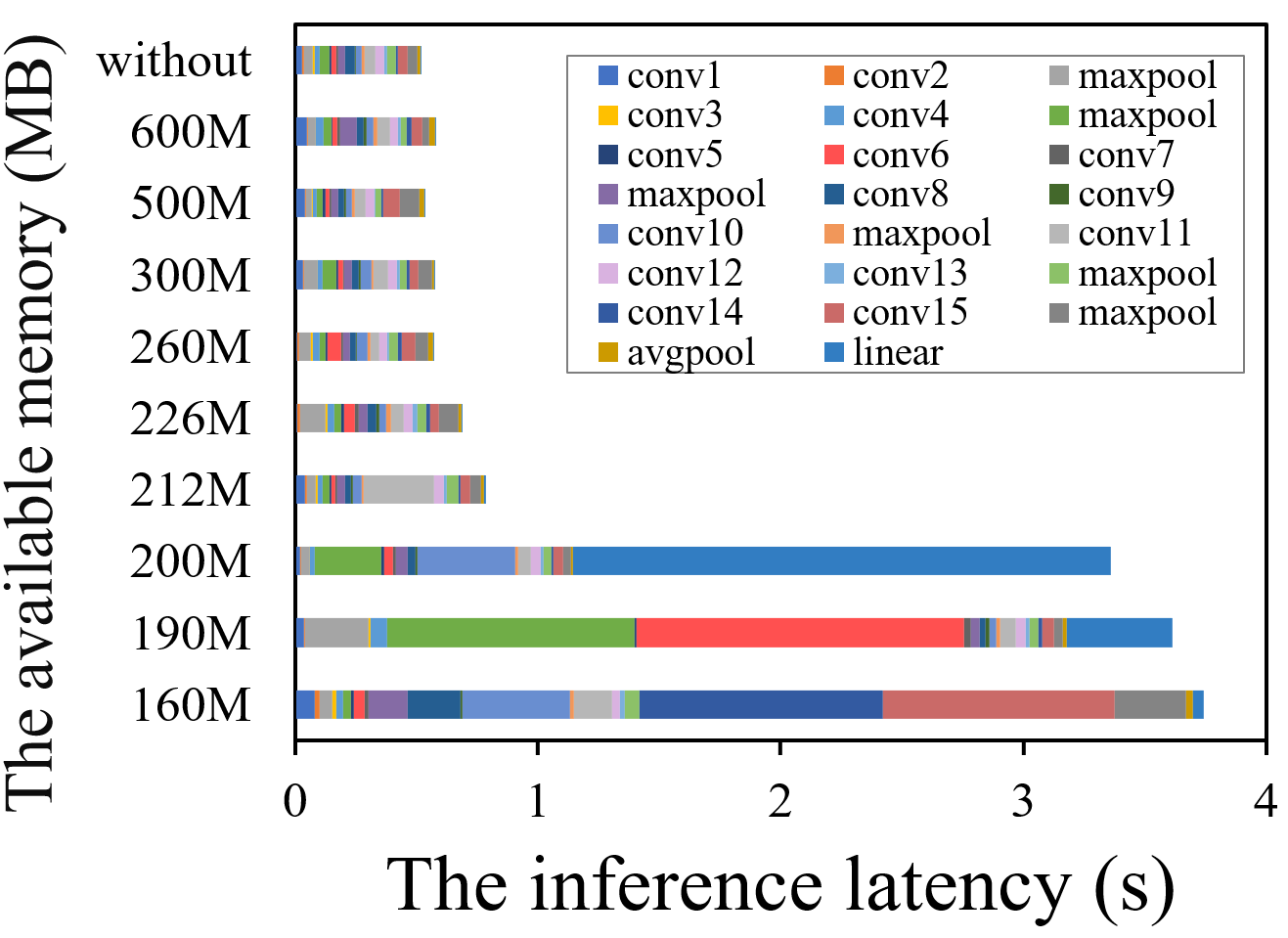}}
  \hfill
  \centering
  \subfloat[GoogLeNet]{
  \includegraphics[height=0.236\textwidth]{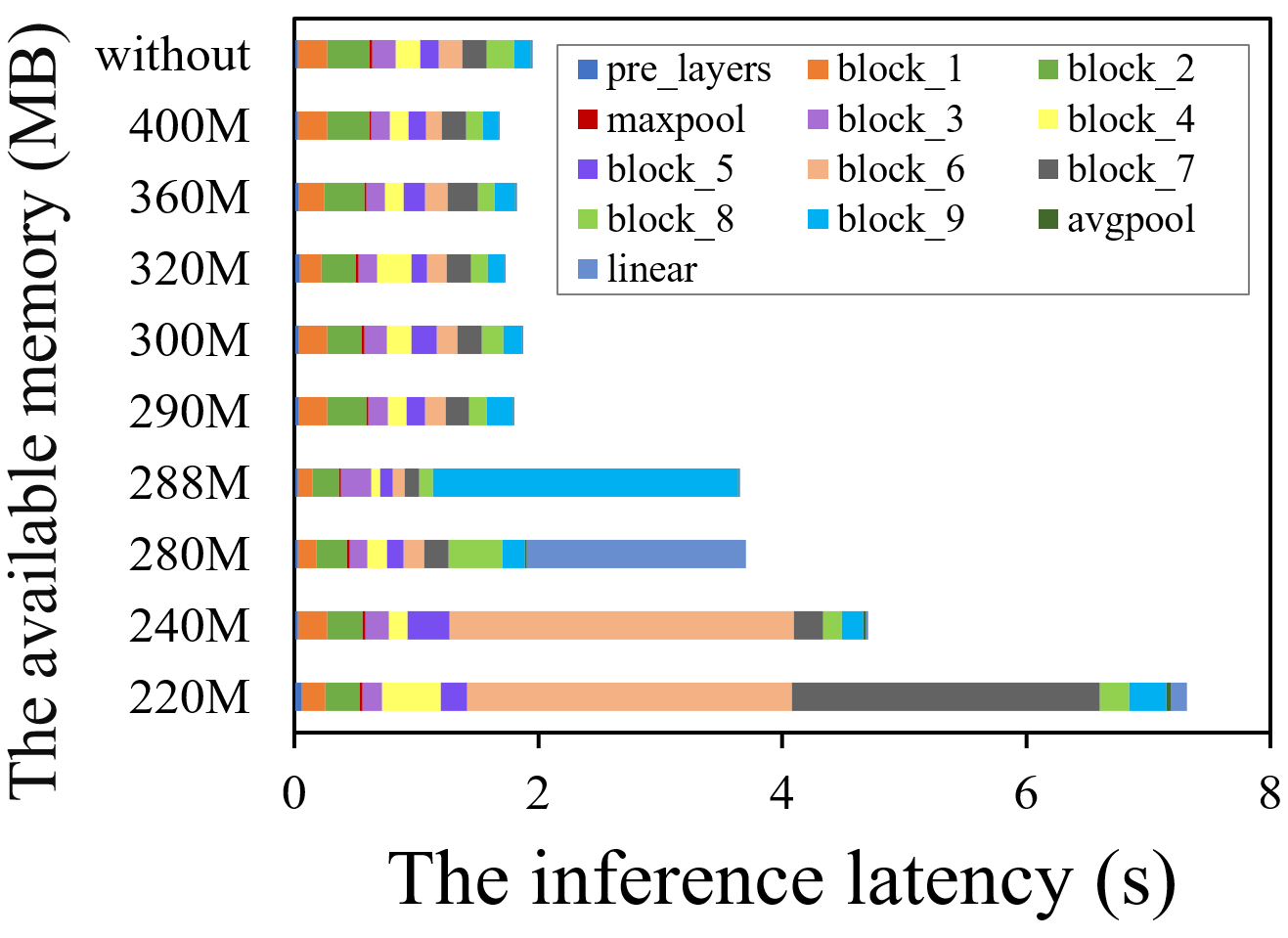}}
\caption{The inference latency of AlexNet, GoogLeNet and VGG16 with changing memory resources on Raspberry Pi 4B. ``without'' means no memory limitation. Different color bars represent the inference latency of different DNN operators. }
\label{img:memory_impact}
\vspace{-1mm}
\end{figure*}

\begin{table*}[t]
\scriptsize
\centering
\caption{Sample spaces design and the number of samples for different DNN operators.}
\label{Sample_space}
\begin{tabular}{|c|c|c|l|c|c|c|}
\cline{1-3} \cline{5-7}
\textbf{Variables} & \textbf{Description} & \textbf{Sample ranges} &  & \textbf{Operator type} & \textbf{Sample variables} & \textbf{ The number of samples} \\ \cline{1-3} \cline{5-7} 
$hw$ & input height and width & [1,512] &  & Conv & $hw$,$k_s$,$cin$,$cout$,$s$ & 12799 \\ \cline{1-3} \cline{5-7} 
$cin$ & input channel number & [1,512] &  & FC & $cin$,$cout$ & 121 \\ \cline{1-3} \cline{5-7} 
$cout$ & output channel number & [1,512] &  & BN & $hw$,$cin$ & 464 \\ \cline{1-3} \cline{5-7} 
$k_s$ & kernel size & $\{$1,3,5,7$\}$ &  & Maxpool & $hw$,$cin$,$k_s$,$s$ & 960 \\ \cline{1-3} \cline{5-7} 
$s$ & stride & $\{$1,2,3$\}$ &  & Avgpool & $hw$,$cin$,$k_s$,$s$ & 960 \\ \cline{1-3} \cline{5-7} 
\end{tabular}
\vspace{-3mm}
\end{table*}

To solve the above challenges, we present to granularly characterize critical impact factors of DNN operators that influence the execution latency to construct the data sampling space. 
Taking Conv as an example, the height and width of the feature map $hw$, the convolutional kernel size $k_s$, the input and output channel number $cin$ and $cout$, and the stride $s$ determine the computation and memory consumption of Conv and accordingly affect the inference latency. Besides, the padding hyperparameters are usually much smaller than the input height or width and can be ignored to simplify the sampling space. 
For each possible DNN configuration hyperparameter, we sample it within the commonly designed range. For example, the sample range of the kernel size $k_s \in \{1,3,5,7\}$ and the stride $s \in \{1,2,3\}$. 
Similarly, we choose important impact factors and determine the sample ranges for other operators, including FC, BN, Maxpool, and Avgpool, to form the final data sample space, as summarized in Table~\ref{Sample_space}. 


\subsection{Context-aware Latency Prediction}
For the latency predictor to function more accurately at runtime, we comprehensively consider the impact of model structural parameters, and dynamic context.

\subsubsection{Impact of Dynamic Memory Resource on Latency}

We experimentally observe that the execution latency will be affected by the running context of deployed devices (\ie available memory resources), resulting in inaccurate latency prediction.
As shown in Fig.~\ref{img:memory_impact}, we use Docker on Raspberry Pi 4B to simulate the dynamic memory change during model running. And we prohibit the memory swap for controlling the maximum available memory resource in the container. The execution latency of different operators in three typical DNNs, including AlexNet, GoogLeNet, and VGG16, are measured with varying memory resources.

Taking GoogLenet as an example, if the available memory $m$ is sufficient (\ie $m\ge 290MB$), the latency remains almost constant when the available memory changes, as shown in the top 5 rows in Fig.~\ref{img:memory_impact}(c). However, after the available memory decreases to a certain threshold, the latency of some operators will increase sharply (\eg the latency of block9 when $m=288MB$, the latency of block6 when $M=240MB$). More seriously, minor shifts in memory will cause significant fluctuations in latency.
This phenomenon still exists in other common DNNs, as shown in Fig.~\ref{img:memory_impact}(a)(b). 
In summary, there is a running memory threshold $M_{0}$ for DNNs. When the available memory resource $m$ of the specific device is greater than $M_{0}$, the execution latency of the DNN is rarely affected by memory changes. In contrast, the model execution latency will increase suddenly when $m<M_{0}$. Consequently, the latency predictor should dynamically revise the predicted latency according to the available memory resources at runtime.

\begin{figure}[t]
    \centering
    \includegraphics[scale=0.33]{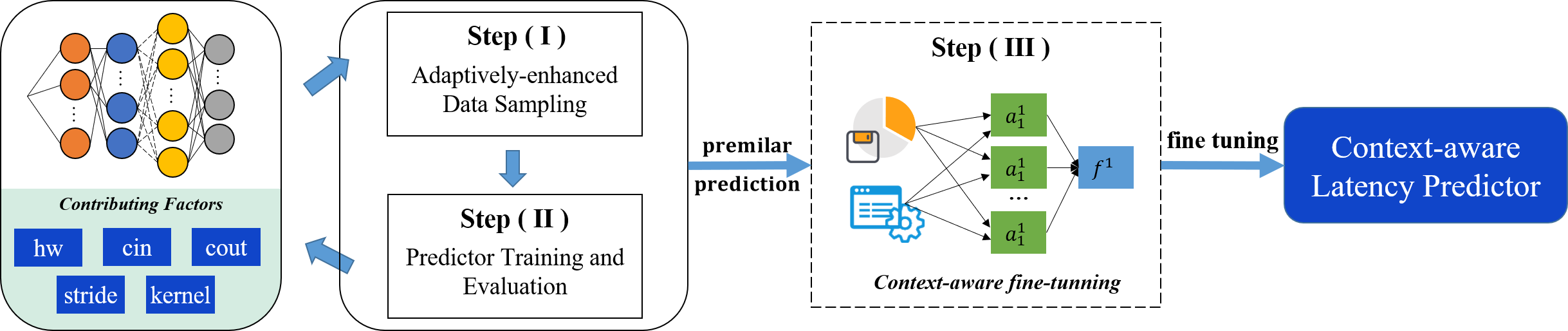}
    \caption{Construction process of the context-aware latency predictor. The accuracy of latency prediction is ensured by considering both model configuration parameters and the dynamic running context of devices. 
    }
    \label{img:latency_predictor}
\end{figure}

\subsubsection{Construction of Context-aware Latency Predictor}
The construction process of the context-aware latency predictor includes three steps, \ie the adaptively-enhanced data sampling, the training and evaluation of random forests-based predictor, and the context-aware fine-tuning by the memory bias, as shown in Fig.~\ref{img:latency_predictor}. 

Based on the elaborately designed data sampling space, we sample enormous amounts of data pairs (\ie [(DNN hyperparameter configurations, latency)]). Further, according to \cite{zhang2021nn-meter}, we design the adaptive supplementary sampling method to add new sampling data for the sampling range where the tested accuracy is lower than the predefined threshold. This process is repeated until the accuracy of the predictor meets the requirements.

To capture the complex non-linear relationships between execution latency and DNN configuration hyperparameters, we select the random forest model \cite{breiman2001random_forests} as the latency prediction backbone, considering its favorable anti-interference ability for abnormal samples.
%
Meanwhile, the weight of each sample and prediction function in the random forest is equal, thus enabling a parallel train of each regressor to reduce the training time. 

Specifically, the random forest-based latency predictor is firstly pre-trained with the adaptively-enhanced sampling data. During the operation of \systemname, it can predict the corresponding execution latency given the atom configuration hyperparameters. Considering the effect of dynamic memory resources, we propose to fine-tune the predicted latency of the specific atom according to the available memory resource of the target device at runtime with a two-layer fully connected model, as shown in Formula \eqref{atom_latency}.



\begin{equation}
T_{p}(atom_i)=\sum_{i=1}^{n}f_{pre}(op_i) + \sum_{i=1}^{n}f_{mem}(op_i, M_{budg})
\label{atom_latency}
\end{equation}
where $\sum_{i=1}^{n}f_{pre}(op_i)$ is the execution latency of operators in $atom_i$ predicted by random forest-based latency predictor $f_{pre}$. $\sum_{i=1}^{n}f_{mem}(op_i, M_{budg})$ is the latency bias produced by the memory budget $M_{budg}$ of the target device.


\begin{equation}
T_p(s_i) = \sum_{i=1}^{m}T_p(atom_i)+T_{tran}(s_i)
\label{model_latency}
\end{equation}
where $T_{tran}$ represents the transmission latency  (see Formula \eqref{equ:tran_latency}), which is determined by the transmission data size and the time-variant network bandwidth.

\section{Performance Evaluation}
\label{sec:experiment}
In this section, we compare \systemnameposs performance with state-of-the-art baselines and present the evaluation of \systemname for diverse DNNs on multiple mobile and edge devices with dynamic deployment context.
Also, the performance of the proposed latency predictor is evaluated over different DNNs and contexts.

\subsection{Experiment Setups}

\begin{table*}[t]
\footnotesize
\centering
\caption{Summary of the datasets and DNNs for evaluating \systemname.}
\vspace{-2mm}
\label{tb:dataset}
\begin{tabular}{|c|c|c|l|c|c|c|c|}
\cline{1-3} \cline{5-8}
\textbf{Dataset}  & \textbf{Task}                     & \textbf{Description} &  & \textbf{Model}             & \textbf{Layer} & \textbf{Model}             & \textbf{Layer} \\ \cline{1-3} \cline{5-8} 
CIFAR100\cite{krizhevsky2009Cifar100} & Image classification     & 60K images  &  & AlexNet \cite{krizhevsky2012alexnet}   & 13    & VGG16 \cite{simonyan2014vgg16}     & 16    \\ \cline{1-3} \cline{5-8} 
ImageNet\cite{deng2009imagenet} & Image classfication      & 14M images  &  & ResNet \cite{li2016resnet101}    & 18    & GoogleNet \cite{szegedy2015googlenet} & 22    \\ \cline{1-3} \cline{5-8} 
BDD100K\cite{yu2018bdd100k}  & Multi-target recognition & 1M videos   &  & MobileNet \cite{howard2017mobilenets} & 28    & Tiny-YOLO \cite{redmon2016tinyyolo} & 26    \\ \cline{1-3} \cline{5-8} 
\end{tabular}
\end{table*}

\textbf{Implementation.}
We implement \systemname on multiple mobile/edge devices with Pytorch \cite{paszke2019pytorch}. 
The mobile device will launch DNN inference tasks to complete classification or object recognition tasks.
And edge devices provide computing and storage resources for computation offloading.
In the offline stage, \systemname conducts the once-for-all DNN pre-partition process and the random forests-based latency predictor training. 
In the online stage, \systemname realizes the context-adaptive DNN atom combination and offloading according to the dynamic deployment context.
When an inference task request arrives, \systemname will establish an asynchronous offloading thread and execution thread, controlled by \textit{DNN Execution Engine}. The \textit{offloading thread} starts to offload the combination of pre-partitioned atoms followed by the offloading order produced by the task initiator (\ie mobile device). 
The \textit{execution thread} executes offloaded atoms on inference task participants (\ie edge devices) and sends an acknowledgment message (ACK) to the task initiator for distributed information synchronization.

\textbf{DNN Models and Mobile Tasks.} 
We select six commonly used DNN models to support diverse latency-sensitive mobile applications (\eg image classification, target recognition) for the evaluation of \systemname, including the small-scale AlexNet \cite{krizhevsky2012alexnet} and VGG16 \cite{simonyan2014vgg16}, the medium-scale ResNet-18 \cite{li2016resnet101} and GoogLeNet \cite{szegedy2015googlenet}, the multi-target recognition model MobileNet \cite{howard2017mobilenets} and Tiny-YOLO \cite{redmon2016tinyyolo}. 
And we conduct experiments on three mobile benchmarks, \ie CIFAR-100\cite{krizhevsky2009Cifar100} and ImageNet\cite{deng2009imagenet} for image classification, and BDD100K\cite{yu2018bdd100k} for multi-target recognition, summarized in Table~\ref{tb:dataset}.

\begin{figure}[t]
  \centering
  \subfloat[Diverse platforms]{
  \includegraphics[height=0.17\textwidth]{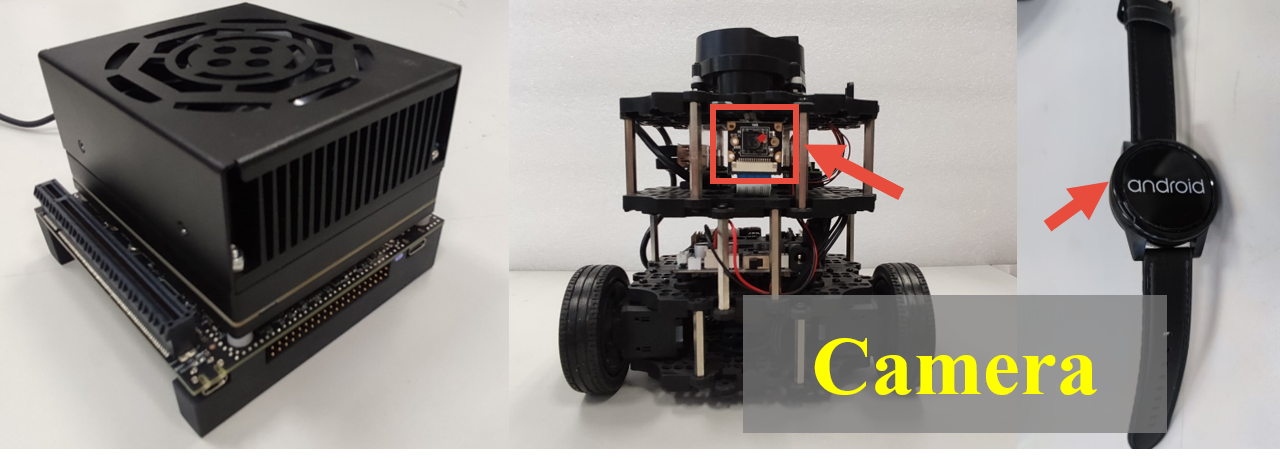}}
  \hfill
  \subfloat[Jetson AGX]{
  \includegraphics[height=0.173\textwidth]{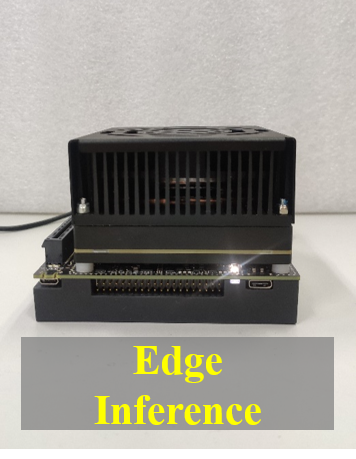}}
  \hfill
  \subfloat[Turtlebot3]{
  \includegraphics[height=0.173\textwidth]{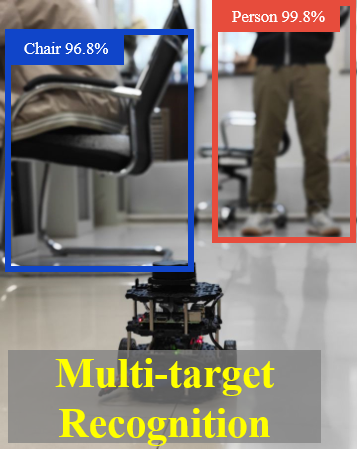}}
  \hfill
  \subfloat[Smart watch]{
  \includegraphics[height=0.173\textwidth]{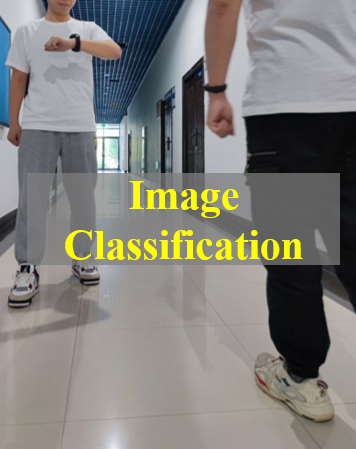}}
\caption{The experimental environment of collaborative computing and DNN inference tasks for \systemname. 
}
\label{differentplatforms}
\vspace{-3mm}
\end{figure}

\textbf{Mobile/Edge Devices with Dynamic Context.}
We evaluated \systemname on three categories of commonly used mobile and edge devices, including the NVIDIA Turtlebot3 based on Raspberry Pi 4B with ARM Cortex-A72 CPU and 2GB memory, the MinDong smartwatch with ARM Cortex-A53 CPU and 3GB memory, and the NVIDIA Jetson AGX Xavier with ARM Cortex-A57 CPU and 8GB memory, as shown in Fig.~\ref{differentplatforms}.
All devices are equipped with WiFi modules and Ethernet ports to establish the network connection. 
The dynamic deployment context is formulated by the latency requirement $T_{user}$, the network bandwidth $B$, and resource budgets (\ie computation budget $C_{bgt}$ and memory budget $M_{bgt}$).

\textbf{Comparison Baselines.}
Firstly, we select seven mobile edge computing methods as baselines to compare with \systemname. 
The details are as below.
%

\begin{itemize}
\item Neurosurgeon \cite{kang2017neurosurgeon} proposes a lightweight scheduler to automatically partition DNN between mobile devices and the server based on the predicted latency and energy consumption.
\item DADS \cite{hu2019dads} transforms the model partition problem into a min-cut problem in the directed acyclic graph, which can optimally partition the DNN under different network conditions.
\item QDMP \cite{zhang2020towards} focuses on reducing the time complexity in finding the optimal partition strategy and proposes a two-stage approach that achieves second-level decision time.
\item CAS \cite{wang2021context} discovers ``the neighbor effect'' to heuristically guide the efficient adaptation of partition solutions under the dynamic running context.
\item IONN \cite{jeong2018ionn} finds the optimal partition points and forms the offloading plan under the guidance of a graph-based heuristic algorithm. The partitioned modules are offloaded to other devices layer by layer.
\item On-device execution executes all DNN layers on the mobile device without offloading. 
\item One-time offloading executes the model only after the offloading process of the whole DNN is completed.
\end{itemize}



Secondly, accurate and timely latency prediction is essential for finding the appropriate DNN partition and combination offloading schemes under dynamic contexts. 
We employ three latency prediction methods to compare with \systemnameposs latency predictor.

\begin{itemize}
\item The linear regression-based prediction (adopted in Neurosurgeon \cite{kang2017neurosurgeon}) fits the correlations between configuration parameters of the deep model and the corresponding inference latency with a linear regression model based on the randomly sampled paired data.
\item The polynomial regression-based (adopted in Edgent \cite{li2019edge}) maps the relationship between model parameters and inference latency with polynomial regression based on the randomly sampled paired data.
\item The random forest-based method fits the correlations based on the randomly sampled paired data, which is the base variant of \systemname's predictor.
\end{itemize}

\subsection{System Performance Comparison}

\begin{table*}[]
\scriptsize
\caption{Performance comparison of \systemname with seven baselines.}
\vspace{-2mm}
\centering
\begin{tabular}{|m{2.3cm}<{\centering}|m{1.5cm}<{\centering}|m{1.7cm}<{\centering}|m{1.7cm}<{\centering}|m{1.5cm}<{\centering}|m{1.9cm}<{\centering}|}
\hline
\textbf{DNN partition techniques} & \textbf{Computation offloading} & \textbf{Collaborative devices} & \textbf{Supported structures} & \textbf{Decision time (ms)} & \textbf{Considered context} \\ \hline
Neurosurgeon \cite{kang2017neurosurgeon}      & N & 2        & Chain     & 428.21 & -                         \\ \hline
DADS \cite{hu2019dads}             & N & 2        & Chain/DAG & 30.01  & Bandwidth                    \\ \hline
QMDP \cite{zhang2020towards}             & N & 2        & Chain/DAG & 32.69  & Bandwidth                    \\ \hline
CAS \cite{wang2021context}              & N & Multiple & Chain     & 2.42   & Bandwidth, energy            \\ \hline
IONN \cite{jeong2018ionn}             & Y & 2        & Chain     & -      & -                         \\ \hline
Once Offload      & Y & 2        & Chain/DAG & -      & -                         \\ \hline
On Scheme Offload & Y & 2        & Chain     & -      & -                         \\ \hline
AdaMEC            & Y & Multiple & Chain/DAG & 2.31   & Bandwidth, energy, and memory \\ \hline
\end{tabular}
\label{tb:performacecompare}
\end{table*}

This subsection presents the evaluation results of \systemname in terms of various performance aspects.

\subsubsection{Comparative Advantages Analysis of \systemname}
To demonstrate the advantages of \systemname in mobile edge computing, we make comparisons with seven baselines from the following five aspects: (1) whether to consider the computation offloading, (2) the maximum number of supported collaborative devices, (3) the support for complex DNN structures, (4) the time consumption for searching a new combination offloading scheme when running context changes and (5) the considered deployment context, as shown in Table~\ref{tb:performacecompare}. All baselines are evaluated on the same inference task (\ie image classification based on CIFAR100) with the same DNN (\ie Alexnet) and mobile devices (\ie Raspberry Pi 4B) for a fair comparison.


\textbf{Comparative analysis.} \systemname outperforms other baselines in various aspects. 
\textit{Firstly}, as mentioned in $\S$ \ref{sec:preliminaries}, it is important to avoid re-partition DNN and deploy the re-partitioned modules when adaptively adjusting partition schemes as the context changes. \systemname prevents extra overheads by decoupling DNN partition and computation offloading, while other baselines, \eg Neurosurgeon, DADS, QMDP, and CAS, need to re-partition and redeploy the model.
\textit{Secondly}, the device-independent DNN pre-partition and combination offloading process ensures that any number of devices can participate in distributed computing to cope with diverse mobile applications. 
As the number of edge devices increases, \systemname will dynamically choose the deployment schemes with the greatest benefit under the guidance of context-adaptive DNN atom combination and offloading algorithm (as shown in \ref{sec:dynamiccontext}).
\textit{Thirdly}, the primitive operator-based DNN pre-partition enables \systemname to support various types of DNNs (\ie chain structure and DAG structure).
\textit{Last but not least}, the \systemnameposs decision time (\ie the time for searching a new partition/offloading scheme when the context changes) is significantly reduced with 0.4$\times$ to 184.4$\times$ improvement than the state-of-the-art baselines over AlexNet. This fast search process enables \systemname to realize the efficient adaptation to dynamic deployment context.
\textit{Finally}, by considering the various deployment contexts (network bandwidth, energy budget, and memory budget), \systemname achieves better flexibility to respond to dynamic context changes.


\begin{figure*}[ht]
  \centering
  \subfloat[Jetson AGX Xavier (AlexNet)]{
  \includegraphics[height=0.22\textwidth]{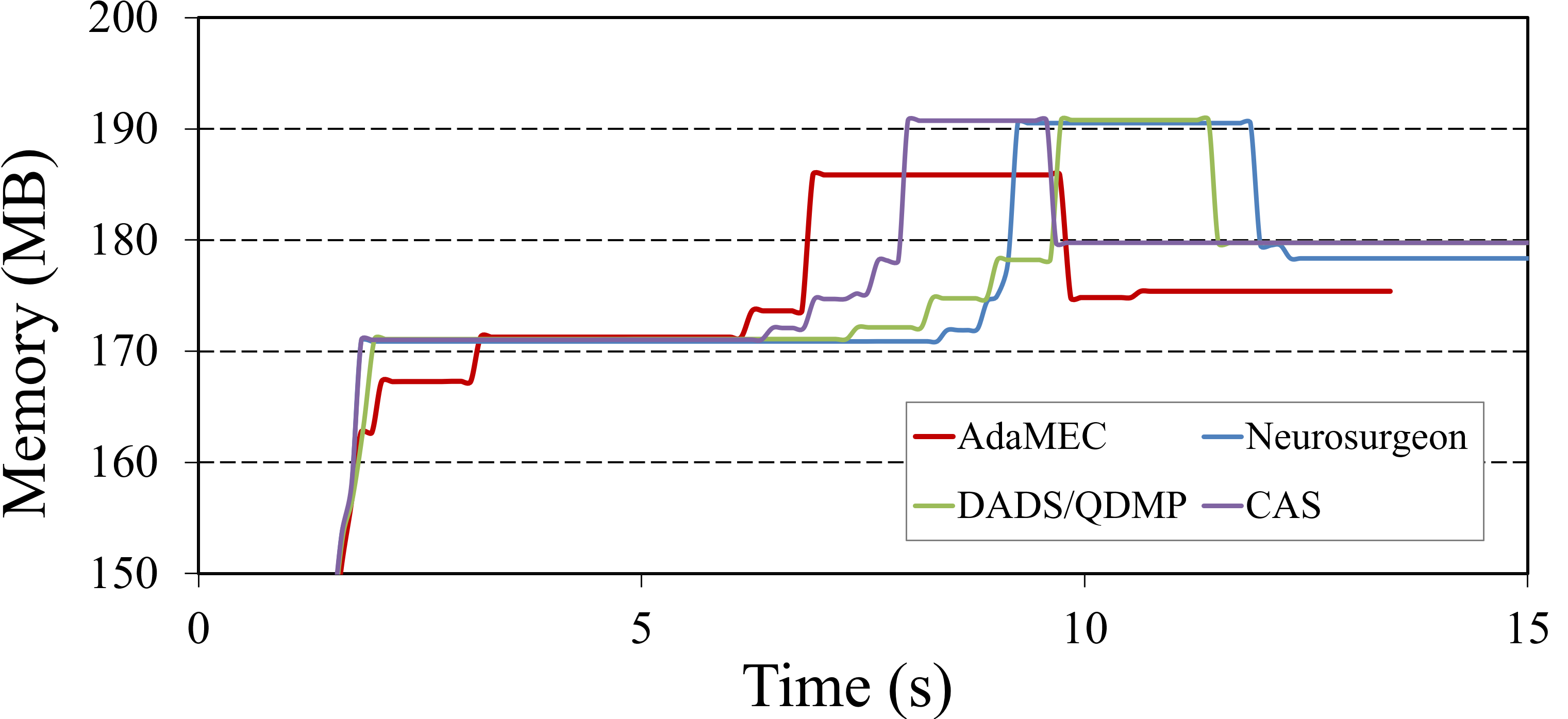}}
  \subfloat[Raspberry Pi 4B (AlexNet)]{
  \includegraphics[height=0.22\textwidth]{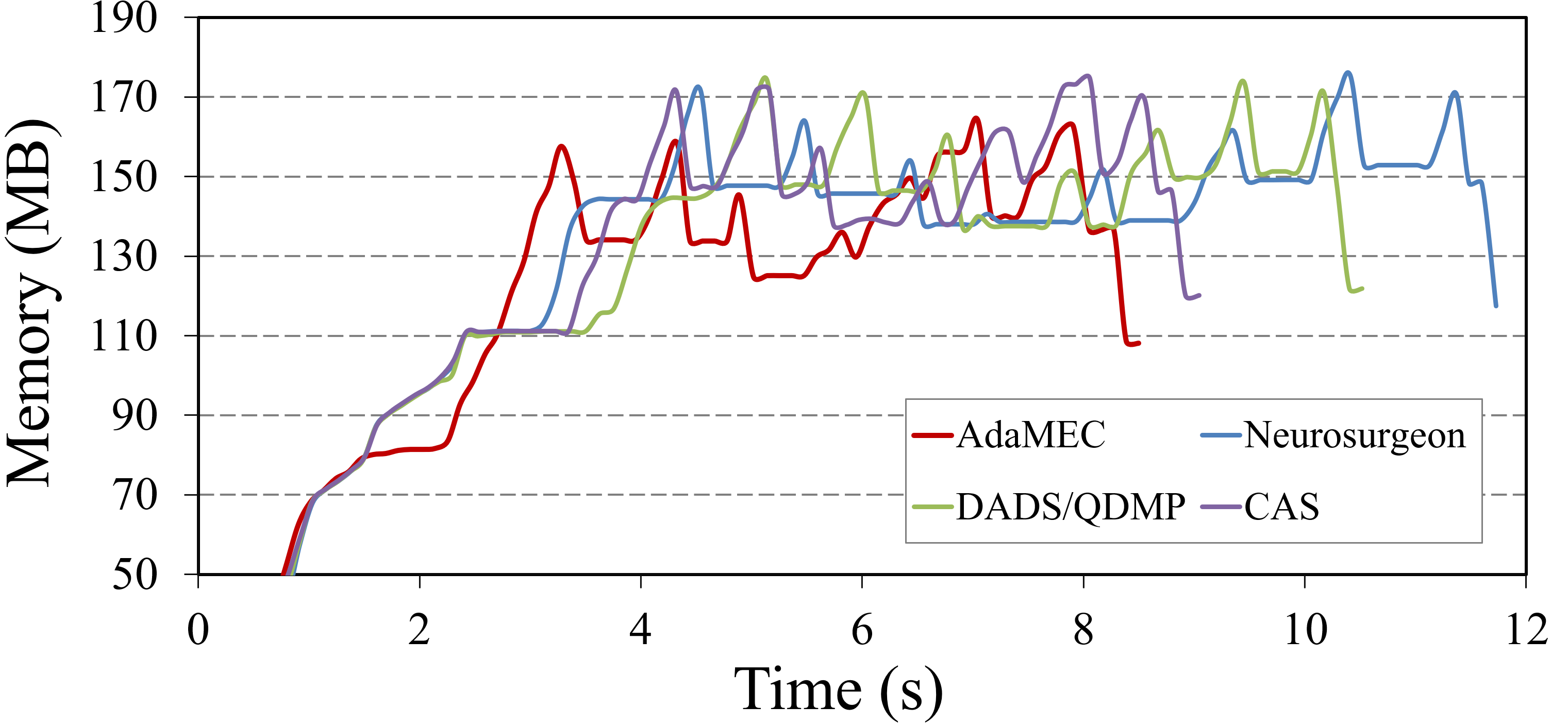}}
  
  \centering
  \subfloat[Jetson AGX Xavier (GoogLeNet)]{
  \includegraphics[height=0.22\textwidth]{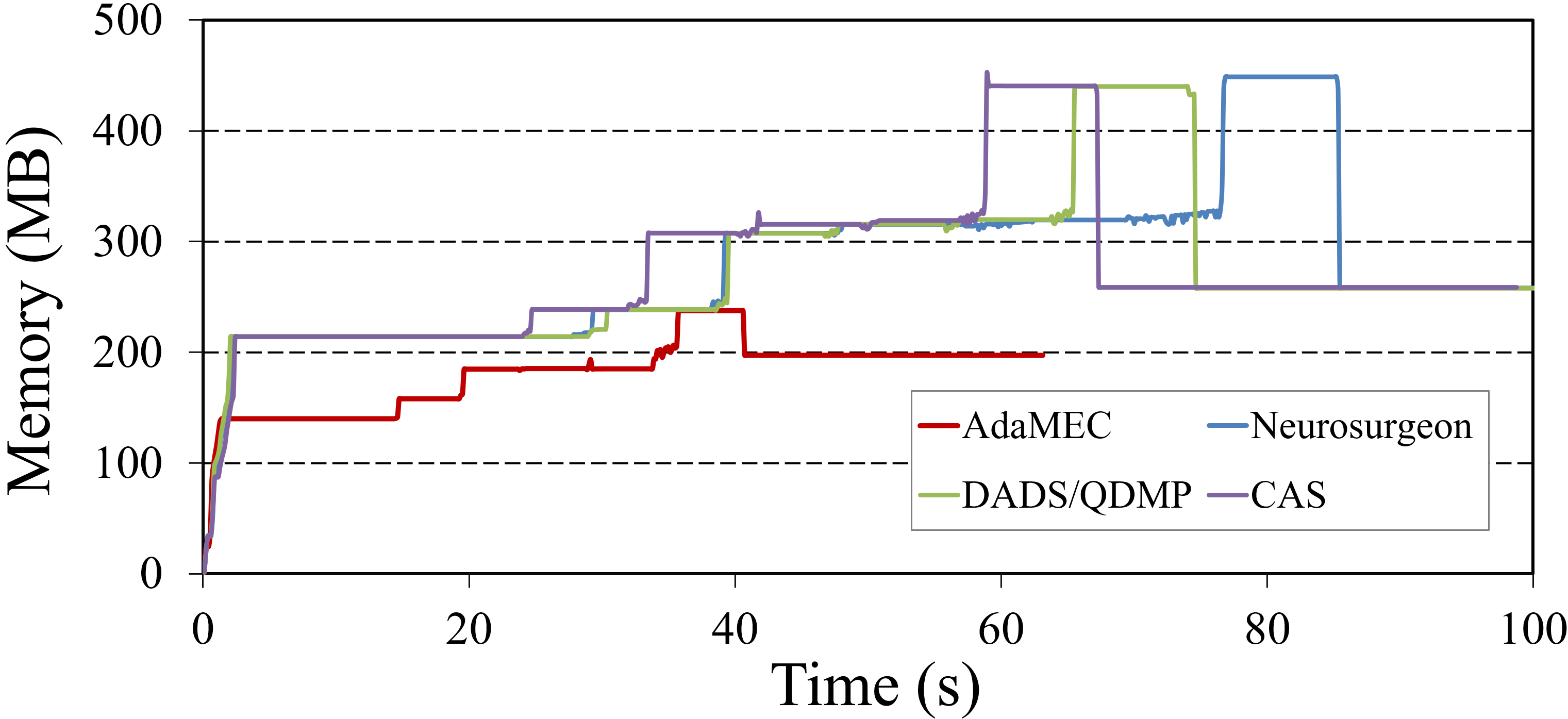}}
  \subfloat[Raspberry Pi 4B (GoogLeNet)]{
  \includegraphics[height=0.22\textwidth]{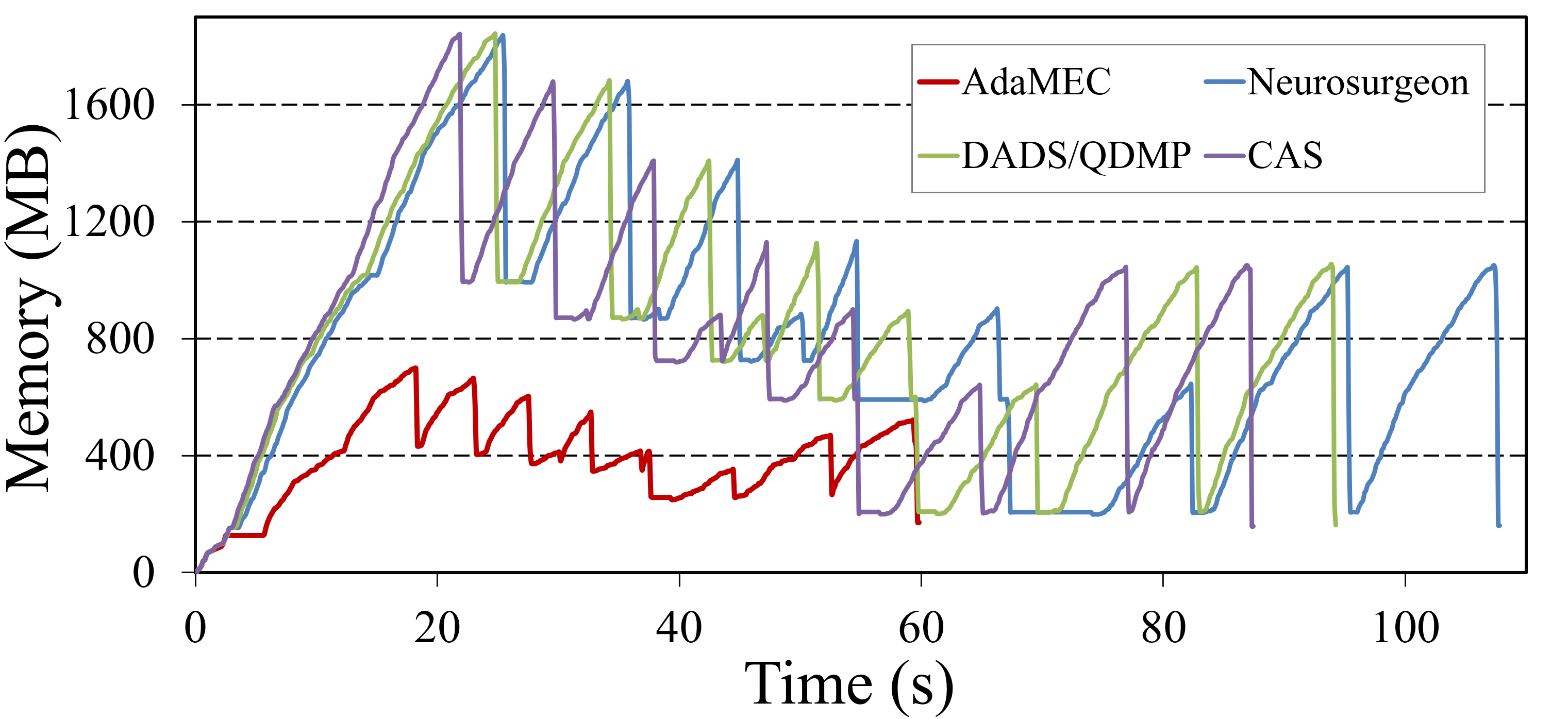}}
\caption{The performance comparison of \systemnameposs once-for-all DNN pre-partition method with baselines on Raspberry Pi 4B and Jetson AGX. 
}
\label{memory_usage_alexnet_and_goglenet}
\vspace{-2mm}
\end{figure*}

\subsubsection{\systemname's performance of once-for-all DNN pre-partition}
To verify the performance of the proposed once-for-all DNN pre-partition block, we compare the memory consumption for DNN inference on different devices during collaborative computing of \systemname and baselines. We adopt the Raspberry Pi 4B as the mobile device and Jetson AGX as the edge device to perform the image classification task based on AlexNet and GoogLeNet, respectively. For fair comparisons, all baselines receive task requests indirectly in the same order and interval. And the network bandwidth, user latency requirements, and resource budgets remain stable and adequate. 

Fig.~\ref{memory_usage_alexnet_and_goglenet} shows the results.
\textit{Firstly}, compared to baselines, \systemname consumes the least memory resources on both mobile and edge devices during the DNN execution process. 
Specifically, on the mobile device, compared with Neurosurgeon, DADS/QDMP, and CAS, the average memory usage of \systemname is reduced by 10.23MB, 12.58MB, and 12.31MB for AlexNet, and 645.88MB, 662.37MB, and 669.43MB for GoogLeNet, respectively. 
And for the edge device, \systemname also realizes superior performance in terms of memory consumption, especially for large-scale DNNs (\eg the average 178.45MB memory savings for GoogleNet, as indicated in Fig.~\ref{memory_usage_alexnet_and_goglenet}(c)). This is because the pre-partition process in \systemname guarantees that only required DNN atoms are offloaded to target devices for DNN execution. 
In this way, \systemname does not need to store the entire model parameters on all edge devices.
This advantage is even more noticeable on mobile devices, which frequently adjust partition schemes, as shown in Fig.~\ref{memory_usage_alexnet_and_goglenet}(b) and Fig.~\ref{memory_usage_alexnet_and_goglenet}(d).  

\textit{Second}, when a new inference task arrives, \systemname will not release the memory occupied by the atoms used by the previous request immediately to save memory usage further. 
When the memory usage exceeds the user memory budget, the retained atoms will be released in the order of first-in-first-out, to avoid the extra memory consumption caused by repeated loading of atoms and ensure that recently executed atoms remain in memory.
\textit{Third}, with the avoidance of repeated calls and loads, \systemname executes task requests faster than other baselines (\ie 48.3\%, 40.6\%, 35\% faster than Neurosurgeon, DADS/QDMP, CAS, respectively) over GoogLeNet.

\textbf{Summary.} \systemname outperforms other baselines in terms of memory consumption and inference latency based on the flexible pre-partitioned DNN atoms, making it ideal for context-adaptive mobile edge computing. 

\begin{figure*}[ht]
  \centering
  \subfloat[AlexNet]{
  \includegraphics[height=0.24\textwidth]{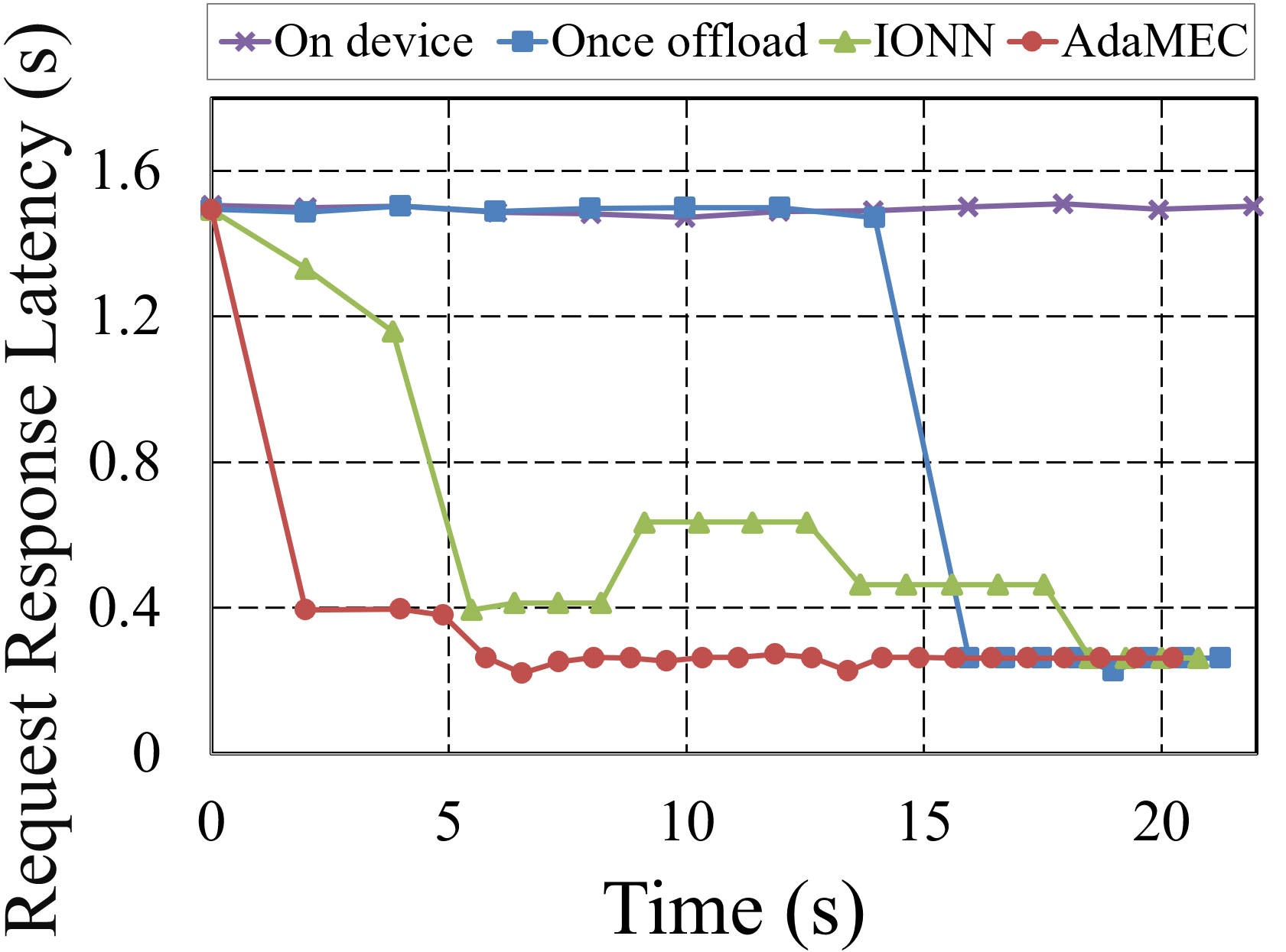}}
  \hfill
  \subfloat[VGG]{
      \includegraphics[height=0.24\textwidth]{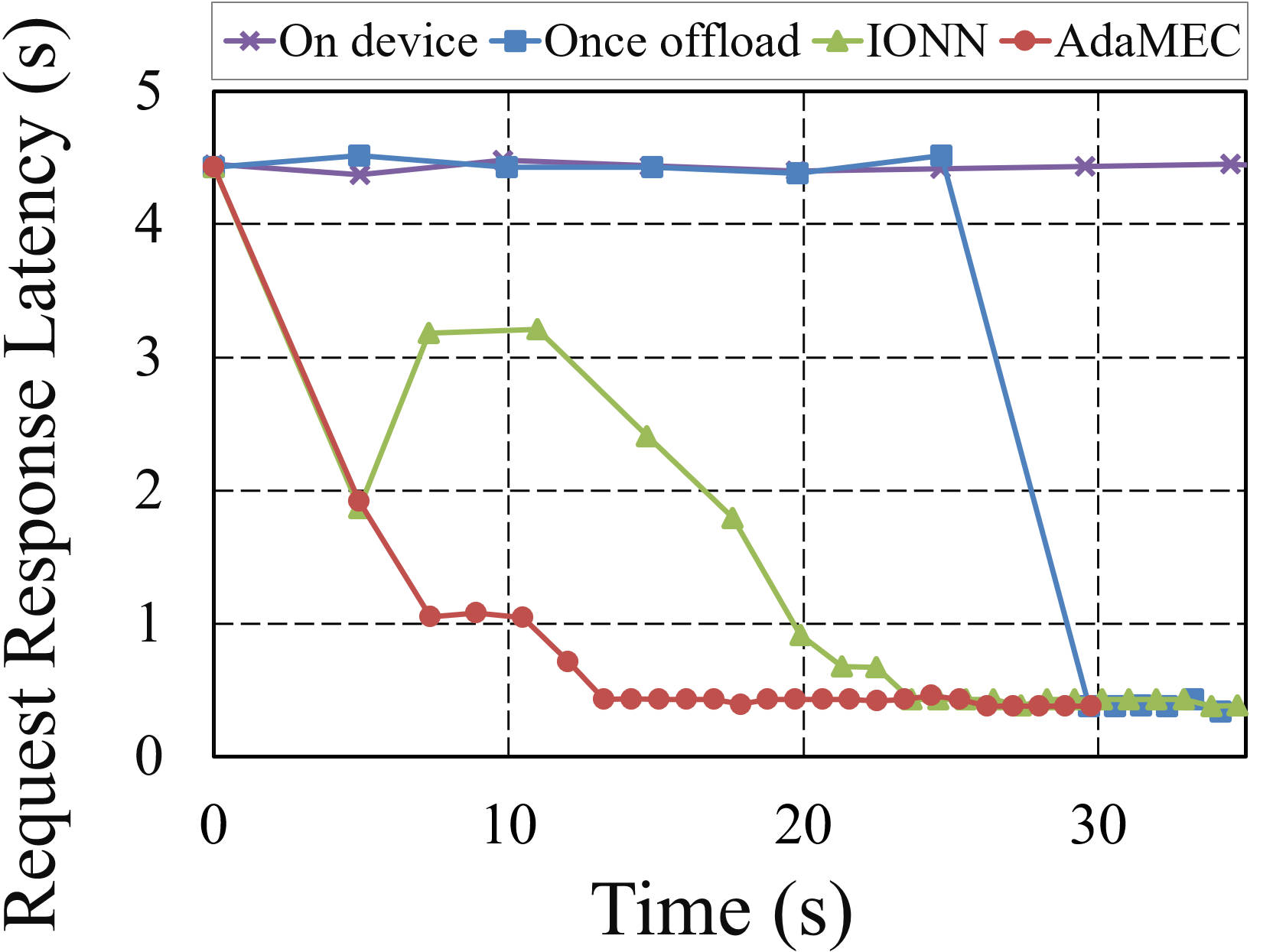}}
  \hfill
  \subfloat[GoogLeNet]{
  \includegraphics[height=0.24\textwidth]{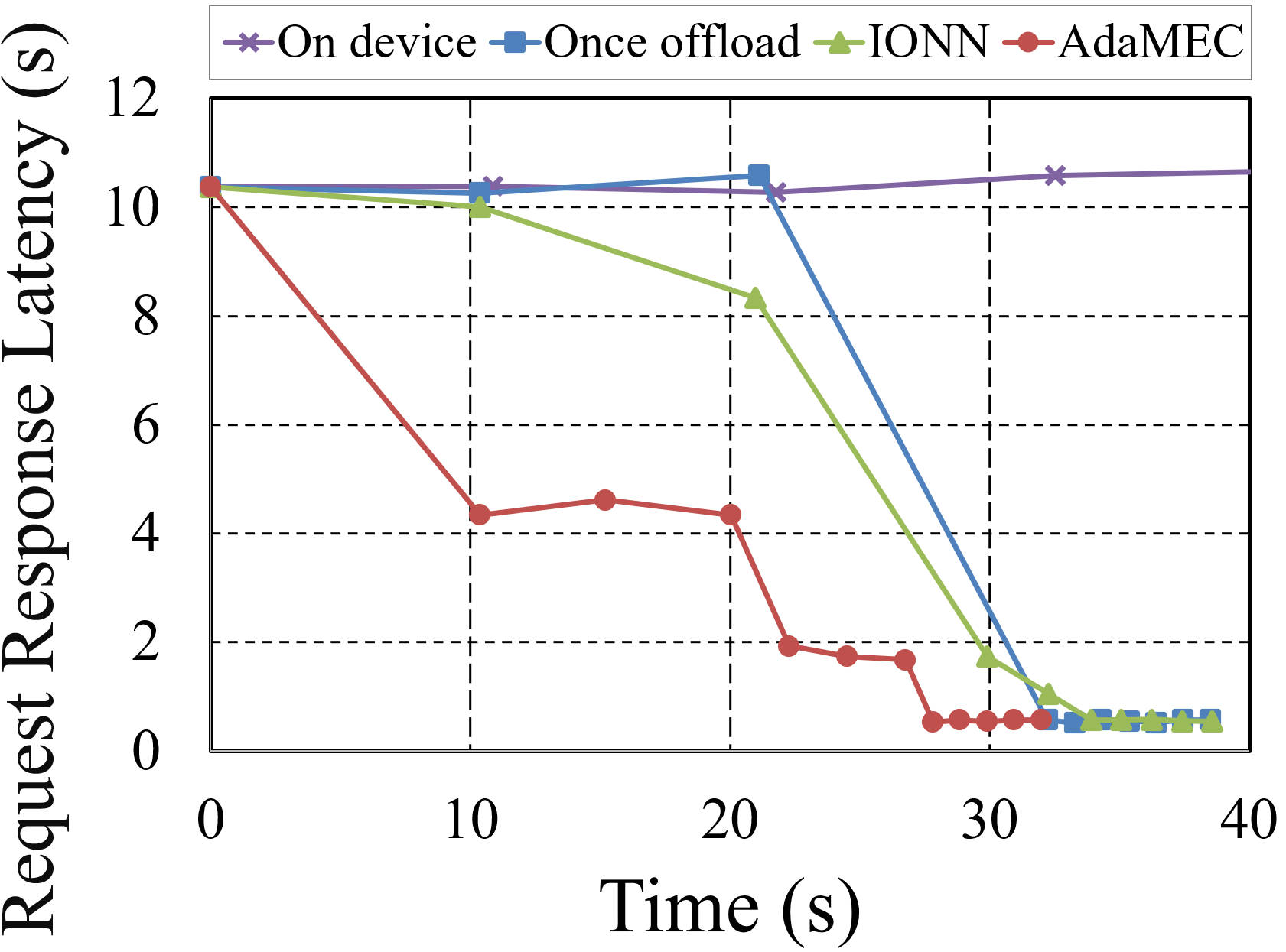}}
  
   \subfloat[ResNet]{
  \includegraphics[height=0.24\textwidth]{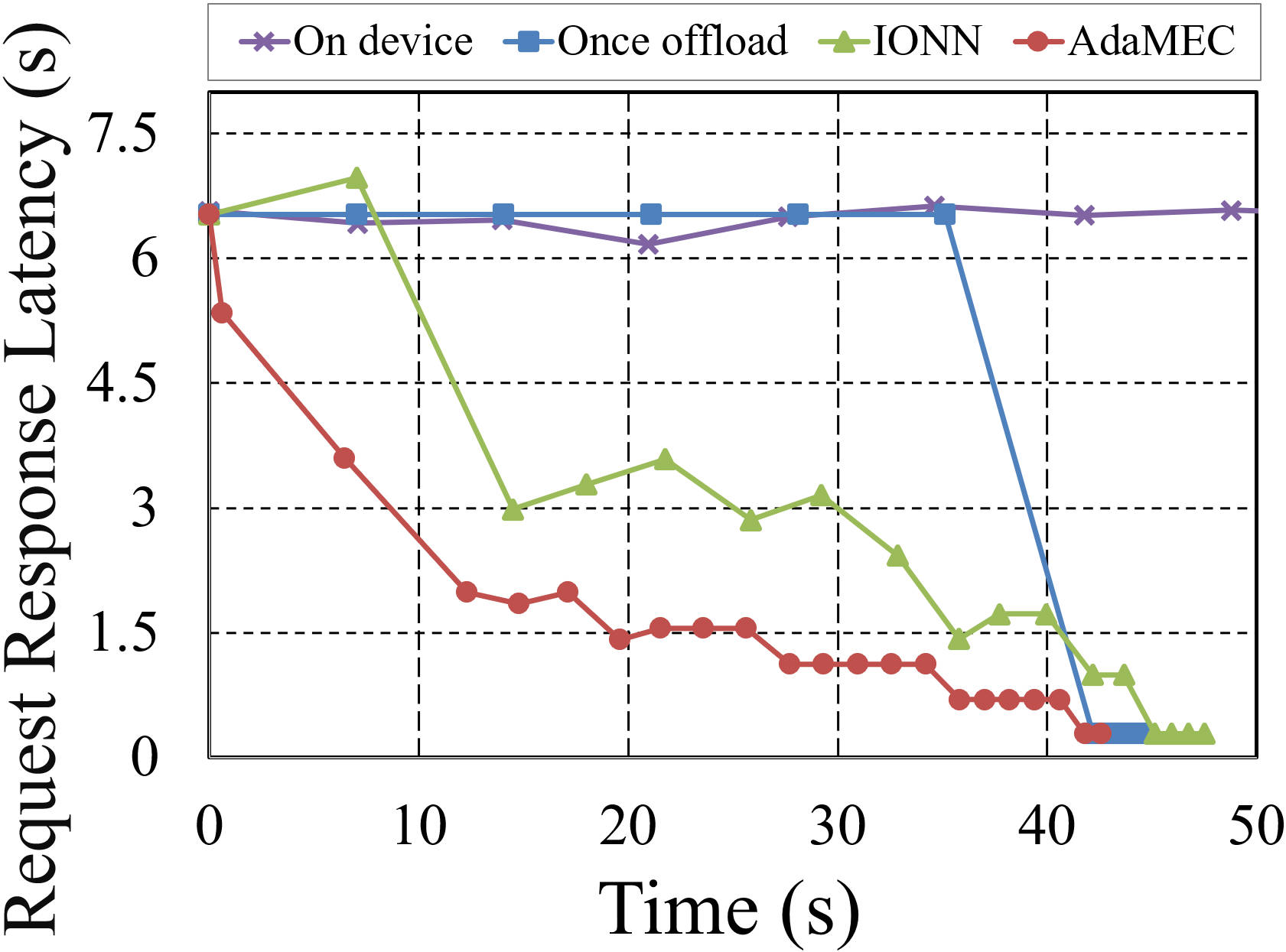}}
  \hfill
  \subfloat[MobileNet]{
  \includegraphics[height=0.24\textwidth]{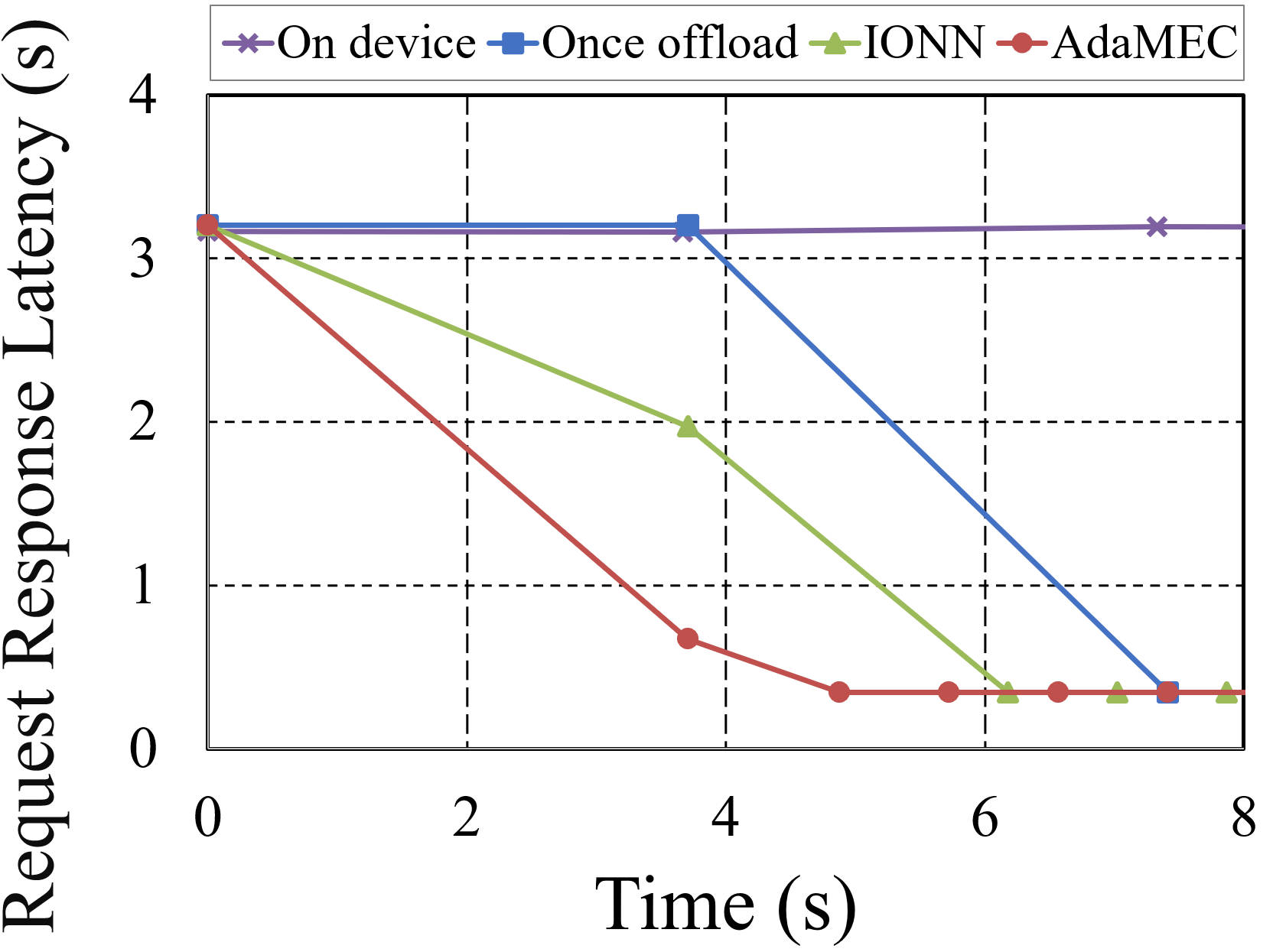}}
  \hfill
  \subfloat[Tiny-YOLO]{
  \includegraphics[height=0.24\textwidth]{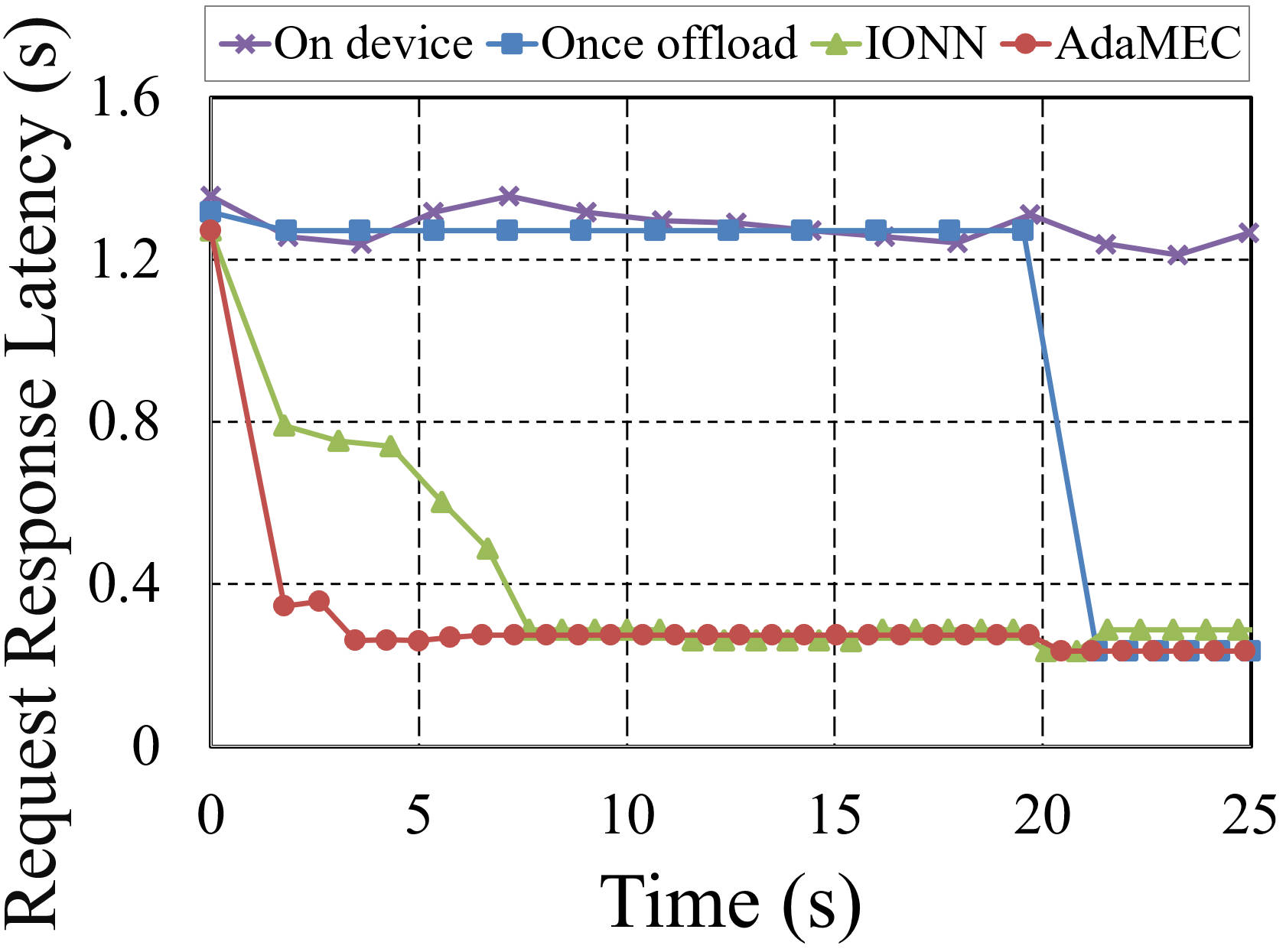}}
\centering
\caption{The performance comparison of \systemnameposs combination offloading with baselines on six typical DNNs. 
}
\label{computation_offloading_performance}
\vspace{-2mm}
\end{figure*}

\subsubsection{\systemname's Performance on context-adaptive DNN atom combination and offloading}
This section evaluates the combinational offloading performance of \systemname by evaluating the response latency of DNN task requests, compared with three baselines considering computation offloading. 
In this thread of experiments, we conduct a comparison of six typical DNNs with different scales for distinct tasks. Raspberry Pi 4B is selected as the mobile device and Jetson AGX serves as the edge device, which are connected via Wi-Fi with an average bandwidth of 40Mbps. 
The latency requirements and resource budgets are adequate and constant for all methods. And the mobile device continuously triggers task requests every 0.5 seconds.

Fig.~\ref{computation_offloading_performance} compares different methods in terms of the response latency of each task request.
The horizontal axis represents the time range of the task request, and the vertical is the response latency.
\textit{Firstly}, the overall response latency of \systemname is lower than all baselines with six DNNs, indicating that the mobile device responds to task requests more quickly from collaborative computing timely. 
Different from the on-device and the once-offloading that need to wait for the complete DNN to be offloaded, \systemname enables edge devices to execute currently offloaded atoms in a timely manner, thus speeding up the collaborative computing process, as shown in Fig.~\ref{computation_offloading_performance}(a).

\textit{Secondly}, \textit{IONN}, the most competitive baseline, which offloads parts of DNN in advance at the layer level, usually results in negative benefits in response latency (\eg the request response latency of IONN is 1.93$\times$ that of \systemname at 6.5s in Fig.~\ref{computation_offloading_performance}(d). This is due to the fact that IONN lacks consideration of the latency benefits of partition points, that is, not every layer offloaded to the edge device will immediately bring faster response latency.
Notably, \systemname prevents such phenomenon by the concept of atoms and combination offloading, which ensures that each atom offloaded to the target device brings a positive latency benefit.
This advantage is more significant for large-scale DNNs, such as ResNet (15$\sim$42s in Fig.~\ref{computation_offloading_performance}(d)) and VGG (5$\sim$10s in Fig.~\ref{computation_offloading_performance}(b)).
\textit{Thirdly}, as shown in Fig.~\ref{computation_offloading_performance}(a)(f), \systemname prioritizes offloading atoms with lower overheads to timely utilize the computing capabilities of edge devices, due to the design of the optimal offloading plan. 
And \systemname reaches a similar latency using less search time, \eg 2.38$\sim$10.15s, 7.25$\sim$16.52s, 4.7$\sim$8.53s, 3.17$\sim$13.84s, 2.22$\sim$19.54s, and 0.85$\sim$2.64s faster on AlexNet, VGG, GoogLeNet, ResNet, Tiny-YOLO, and MobileNet tasks, respectively.

\textbf{Summary.} Based on the context-adaptive DNN atom combination and offloading, \systemname demonstrates the advantages in the response latency with mobile edge computing. 
And \systemname also takes the least amount of search time to obtain the optimal offloading benefits.

\begin{figure*}[t]
  \centering
  \subfloat[Scenario A]{
  \includegraphics[height=0.23\textwidth]{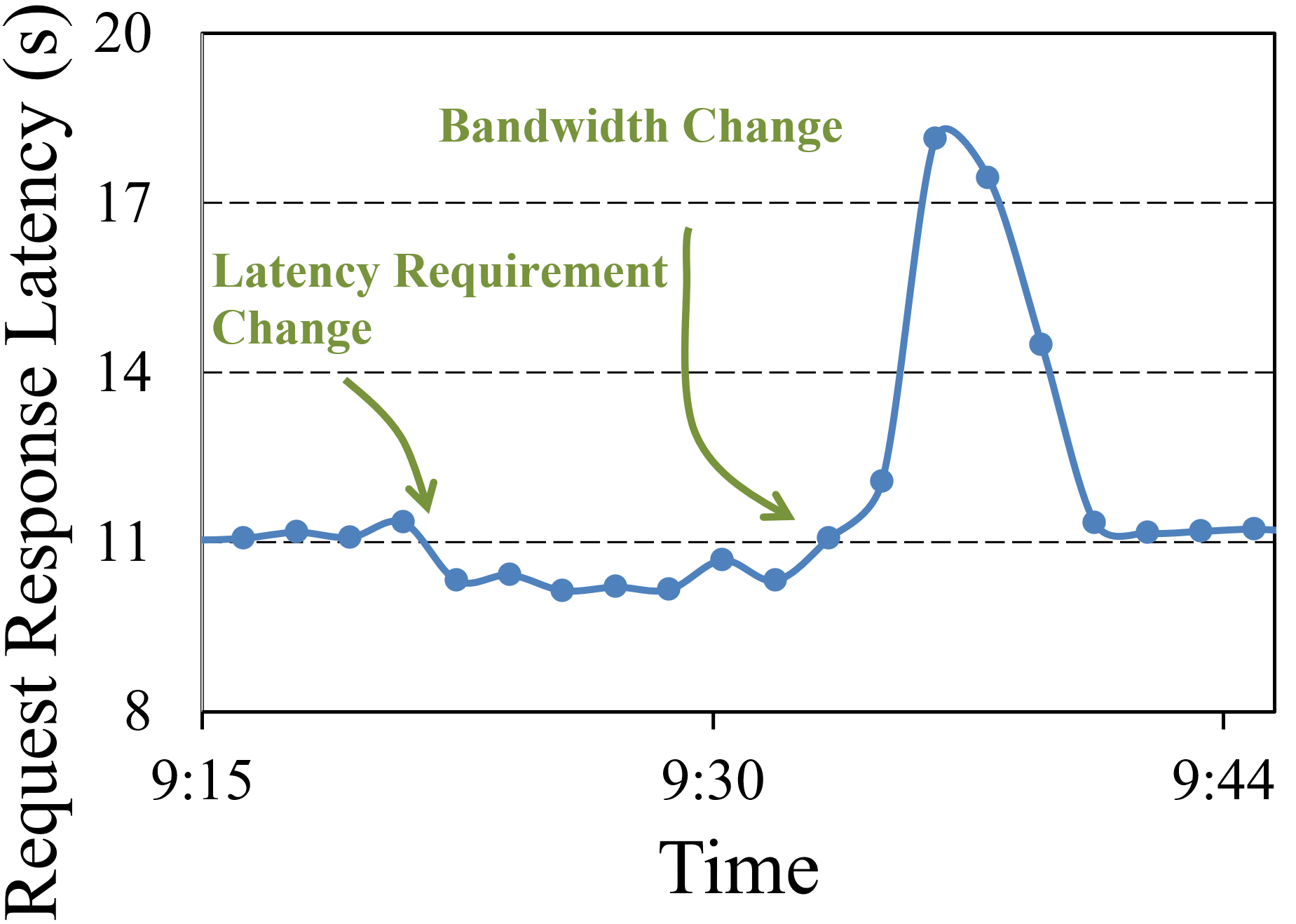}}
  \hfill
  \subfloat[Scenario B]{
  \includegraphics[height=0.23\textwidth]{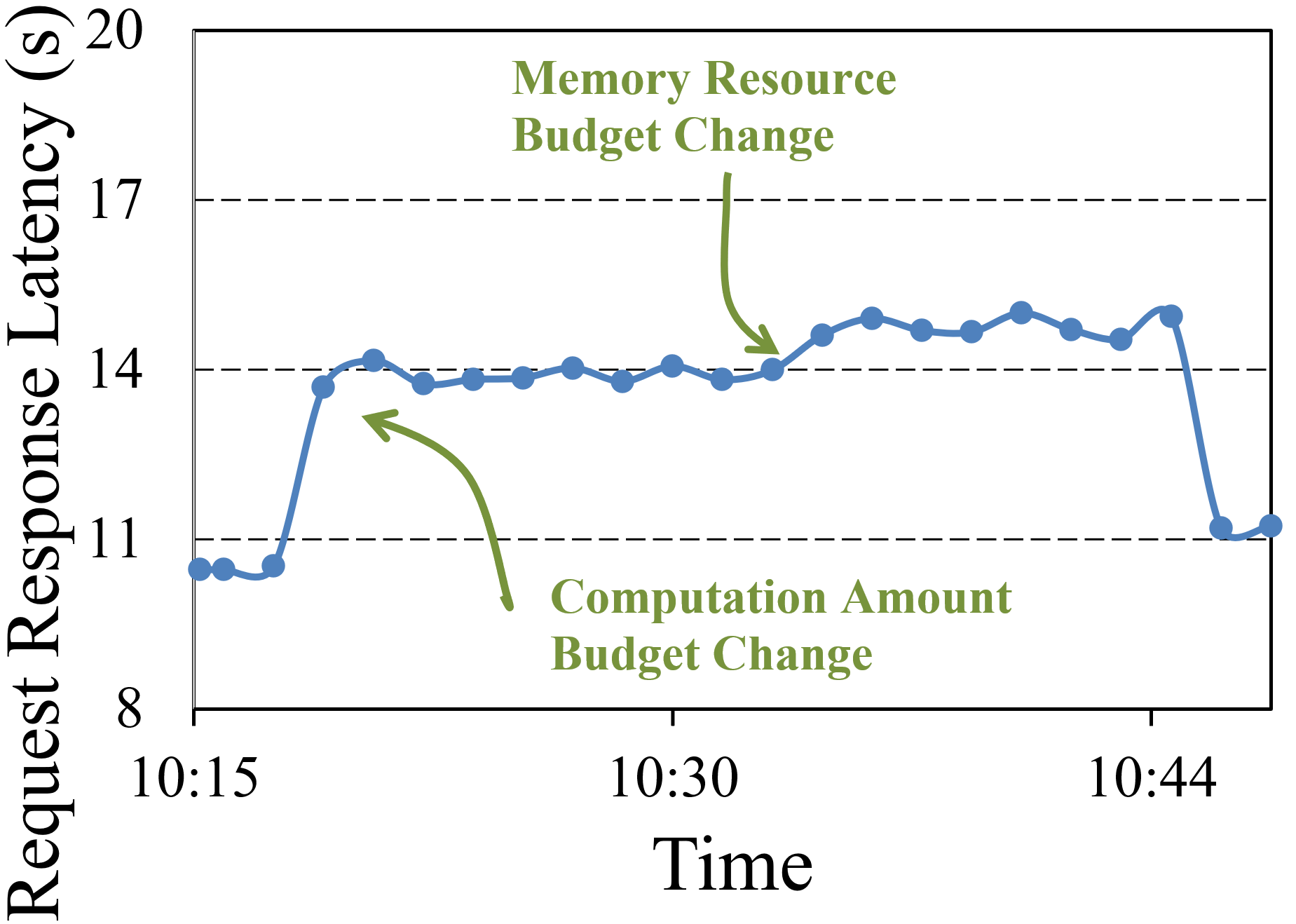}}
  \hfill
  \subfloat[Scenario C]{
  \includegraphics[height=0.23\textwidth]{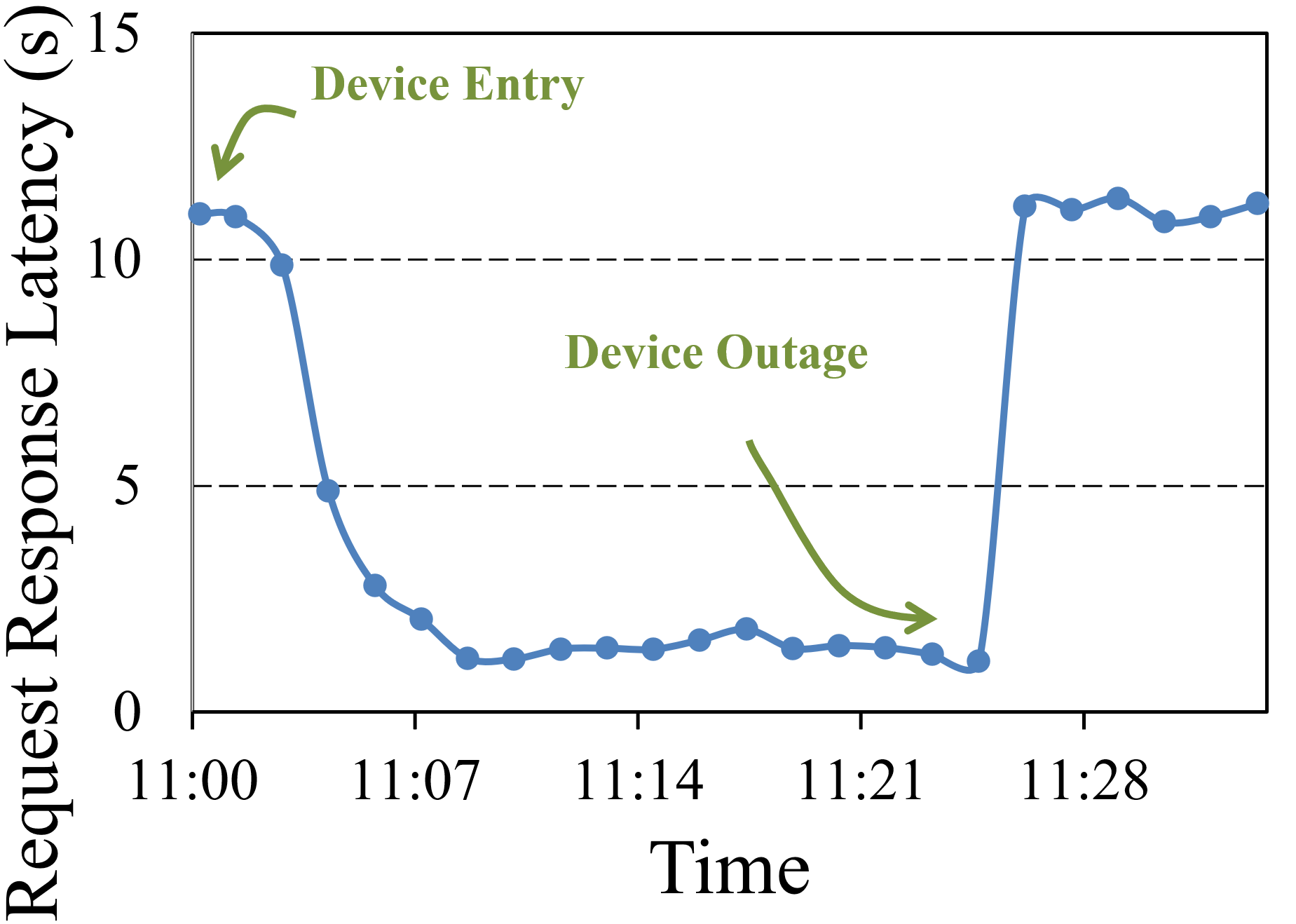}}
  
  \centering
  \subfloat[Smart Watch]{
  \includegraphics[height=0.23\textwidth]{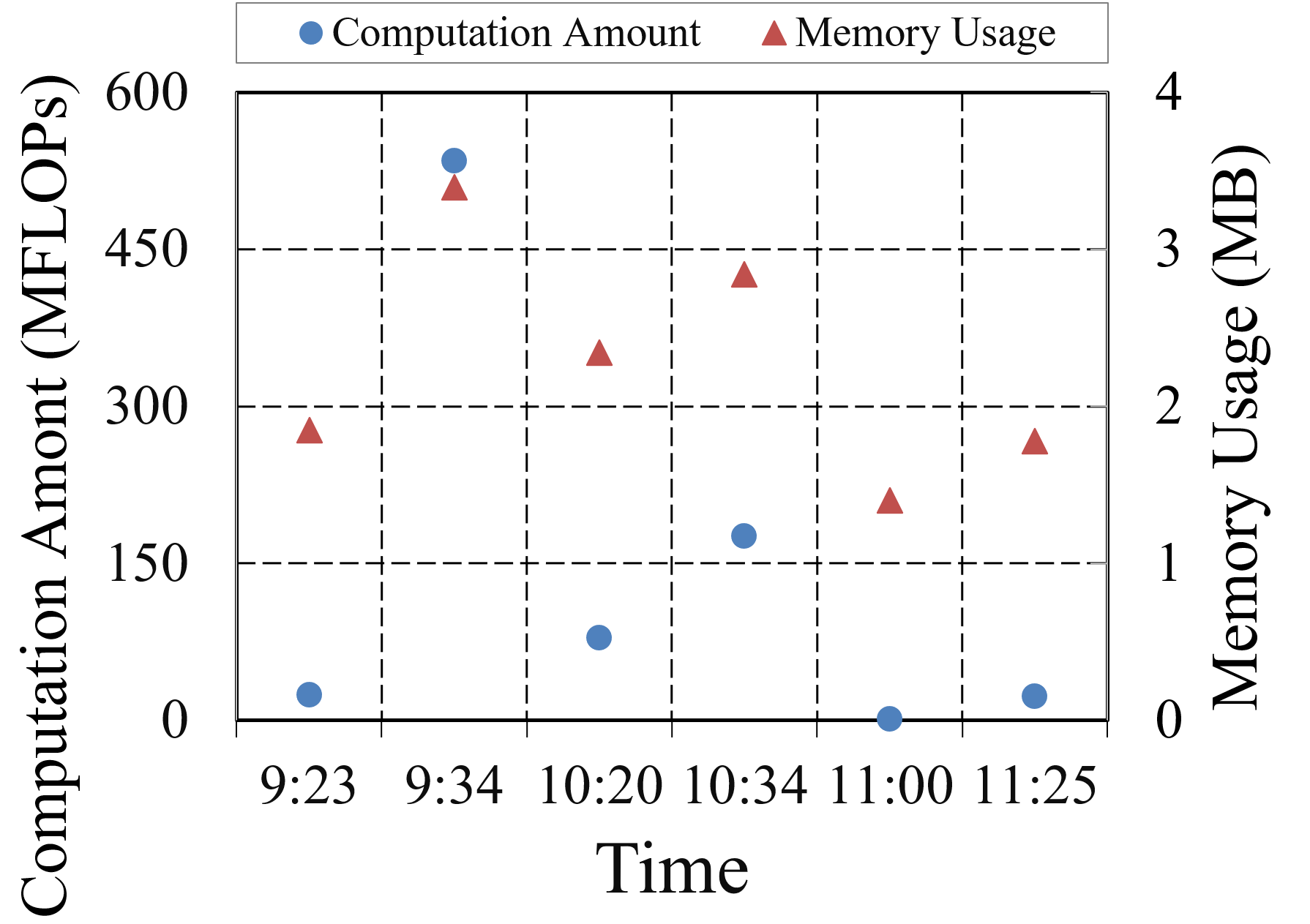}}
  \hfill
  \subfloat[Raspberry4B]{
  \includegraphics[height=0.23\textwidth]{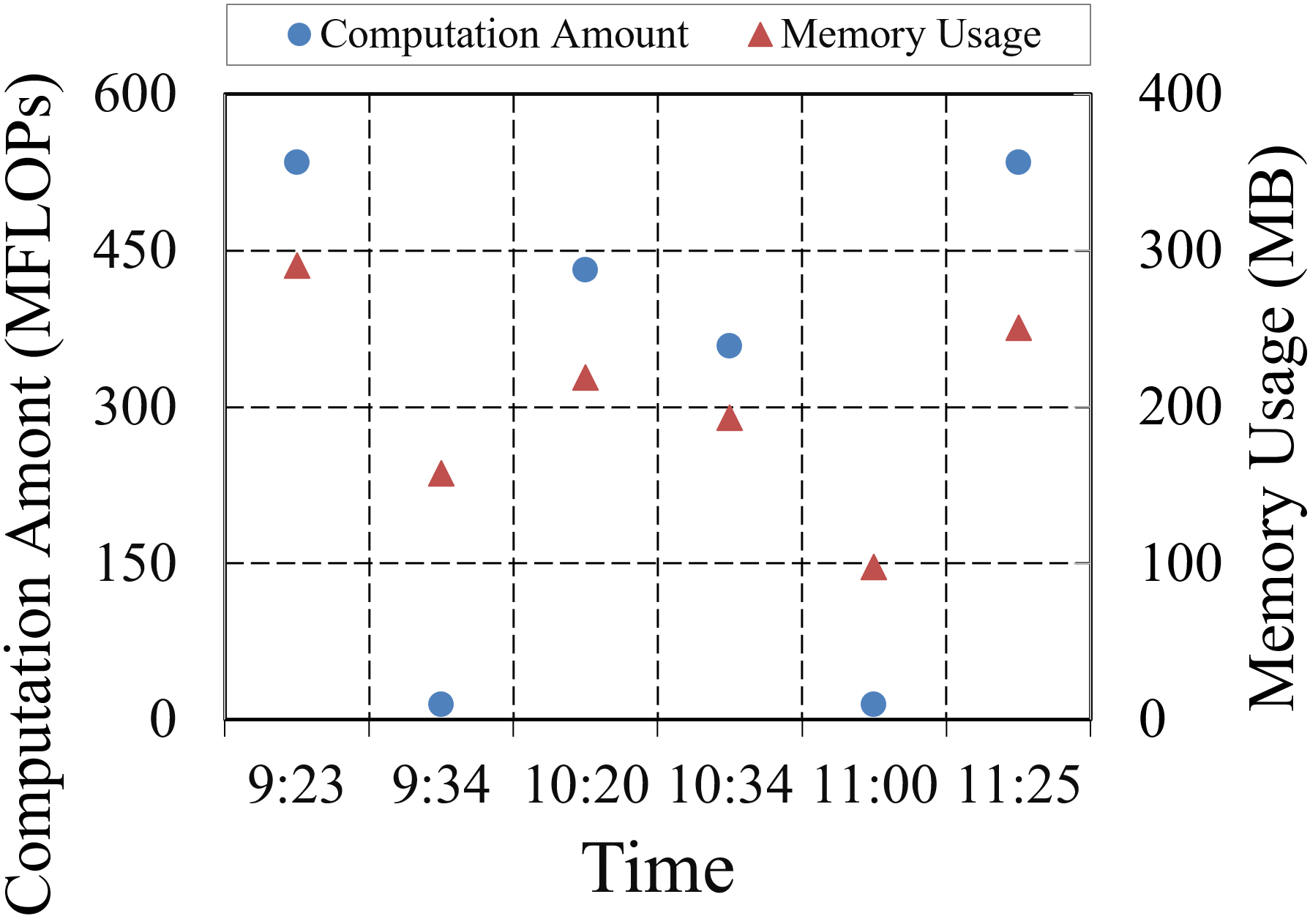}}
  \hfill
  \subfloat[Nvidia Jetson AGX]{
  \includegraphics[height=0.23\textwidth]{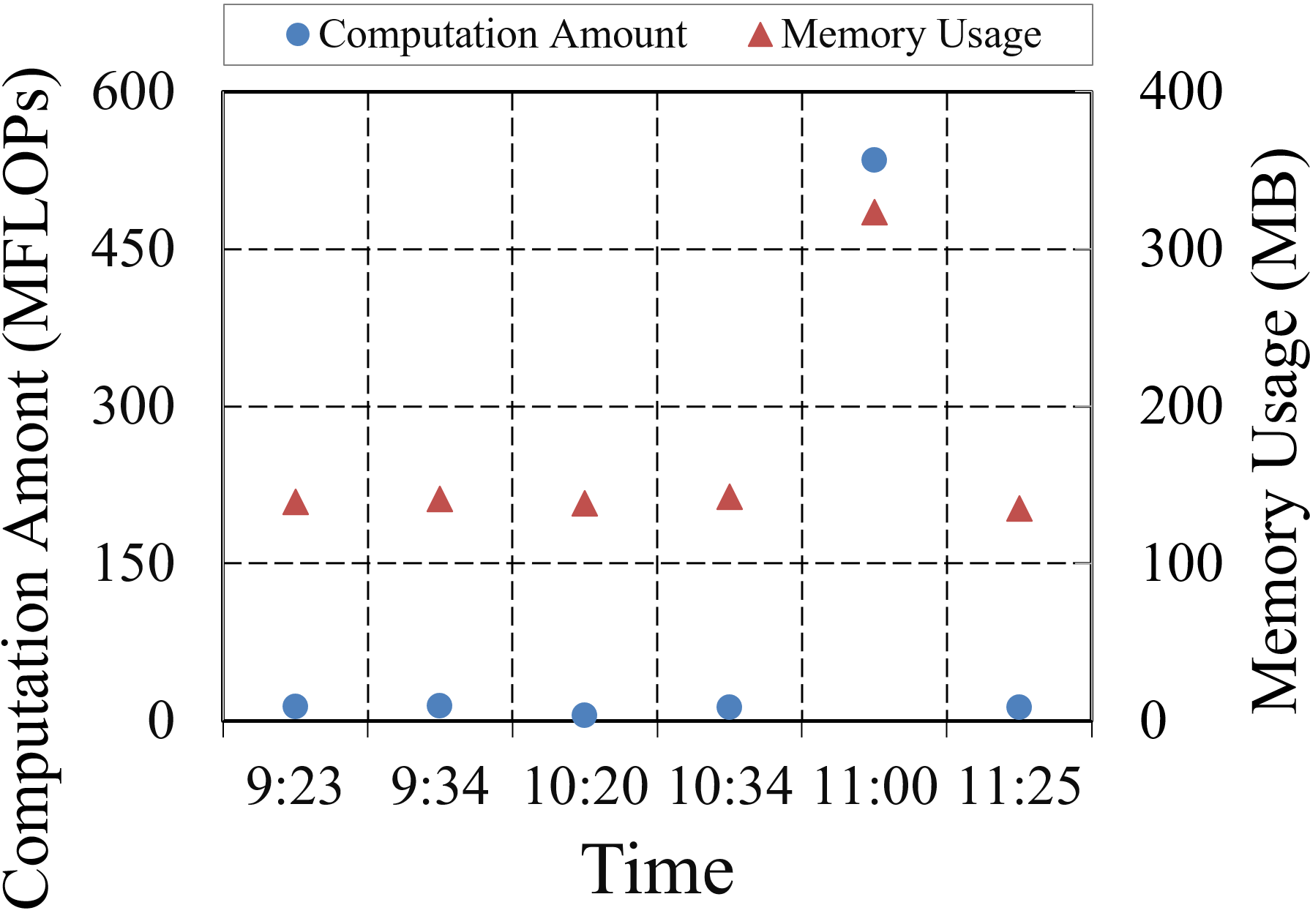}}
\caption{The runtime and adaptive combination offloading process of \systemname under the dynamic contexts. 
}
\label{adaptive_ability}
\vspace{-2mm}
\end{figure*}

\begin{table*}[t]
\centering
\caption{The dynamic context settings on six specific time points during the continuous running of \systemname.}
\scriptsize
\begin{tabular}{|ccccccc|}
\hline
\multicolumn{7}{|c|}{\textbf{Dynamic contexts}}                                                                                               \\ \hline
\multicolumn{1}{|c|}{\textbf{Time}}                           & \multicolumn{1}{c|}{9:21$am$}    & \multicolumn{1}{c|}{9:36$am$}    & \multicolumn{1}{c|}{10:20$am$}   & \multicolumn{1}{c|}{10:30$am$}   & \multicolumn{1}{c|}{11:00$am$}         & 11:25$am$   \\ \hline
\multicolumn{1}{|l|}{\textbf{Latency requirement $T_{user}$}} & \multicolumn{1}{c|}{10s}         & \multicolumn{1}{c|}{11s}         & \multicolumn{1}{c|}{11s}         & \multicolumn{1}{c|}{11s}         & \multicolumn{1}{c|}{11s}               & 11s         \\ \hline
\multicolumn{1}{|c|}{\textbf{Bandwidth $B$}}                  & \multicolumn{1}{c|}{33.8Mbps}    & \multicolumn{1}{c|}{11.2Mbps}    & \multicolumn{1}{c|}{25.6Mbps}    & \multicolumn{1}{c|}{40.3Mbps}    & \multicolumn{1}{c|}{38.9Mbps}          & 35.5Mbps    \\ \hline
\multicolumn{1}{|c|}{\textbf{Memory budget $M_{budg}$}}       & \multicolumn{1}{c|}{300MB}       & \multicolumn{1}{c|}{300MB}       & \multicolumn{1}{c|}{300MB}       & \multicolumn{1}{c|}{150MB}       & \multicolumn{1}{c|}{300MB}             & 300MB       \\ \hline
\multicolumn{1}{|c|}{\textbf{Computation budget $C_{budg}$}}  & \multicolumn{1}{c|}{550 MFLOPs}  & \multicolumn{1}{c|}{550MFLOPs}   & \multicolumn{1}{c|}{450MFLOPs}   & \multicolumn{1}{c|}{450MFLOPs}   & \multicolumn{1}{c|}{550MFLOPs}         & 550MFLOPs   \\ \hline
\multicolumn{1}{|c|}{\textbf{Participating devices}}          & \multicolumn{1}{c|}{$D_A$,$D_B$} & \multicolumn{1}{c|}{$D_A$,$D_B$} & \multicolumn{1}{c|}{$D_A$,$D_B$} & \multicolumn{1}{c|}{$D_A$,$D_B$} & \multicolumn{1}{c|}{$D_A$,$D_B$,$D_C$} & $D_A$,$D_B$ \\ \hline
\end{tabular}
\label{context_settings}
\end{table*}

\subsubsection{\systemname's Performance across Diverse and Dynamic Deployment Context}
\label{sec:dynamiccontext}
We conduct continuous evaluations with dynamically changing contexts (\eg network conditions, memory budgets) to verify the adaptive capacity of  \systemname in real-world scenarios.
We deploy the image recognition application based on GoogLeNet on three mobile and edge devices, \ie Mingdong Smart Watch ($D_1$) as the mobile device, Raspberry Pi 4B ($D_2$) and NVIDIA Jetson AGX ($D_3$) as edge devices. 
And the DNN application continuously runs from 9:00 $am$ to 12:00 $am$ with diverse and dynamic running contexts.
The other settings of devices are consistent with $\S$ 5.2.2.
We select the following three mobile scenarios to demonstrate the performance of \systemname: (1) \textit{Scenario A}: the network bandwidth and user latency requirements change, (2) \textit{Scenario B}: device memory budgets and computation amount budgets change, (3) \textit{Scenario C}: a new device suddenly entry/outage.
And we pick three time fragments corresponding above scenarios and six moments among them to analyze the adaptive capability of\systemname in dynamic contexts, as shown in Table.~\ref{context_settings}.

Fig.~\ref{adaptive_ability} illustrates the request response latency (\ie DNN inference latency) and the resource status of different devices during the running of \systemname. 
When latency requirement changes, \systemname adjusts the current combination offloading plan to meet it, as shown in Fig.~\ref{adaptive_ability}(a). 
With the gradual decrease of bandwidth at 9:30, the combination offloading plan changes, for reducing the data transmission cost, as reflected by the increase of resources in Fig.~\ref{adaptive_ability}(d).
Similarly, when the computation or the memory budget on device $D_B$ is reduced, \systemname leaves more atoms on the mobile device $D_A$, as shown in Fig.~\ref{adaptive_ability}(b)(d).
Remarkably, when a new device $D_C$ with stronger computing power participates in mobile edge computing, \systemname offloads more atoms to the new device to accelerate the DNN inference, as shown in Fig.~\ref{adaptive_ability}(c)(f).
Moreover, when a participating device is disconnected, \systemname re-offloads atoms on it to available devices to ensure the smooth DNN inference, as shown in Fig.~\ref{adaptive_ability}(c)(e).

\textbf{Summary.} \systemname adaptively makes DNN atoms combination and offloading to improve the mobile edge computing performance under dynamic contexts.


\subsection{Performance of Runtime Latency Predictor}
To illustrate the performance of \systemnameposs runtime latency predictor, we evaluate it over different DNNs and dynamic deployment contexts.

\begin{figure}[t]
  \centering
  \includegraphics[height=0.3\textwidth]{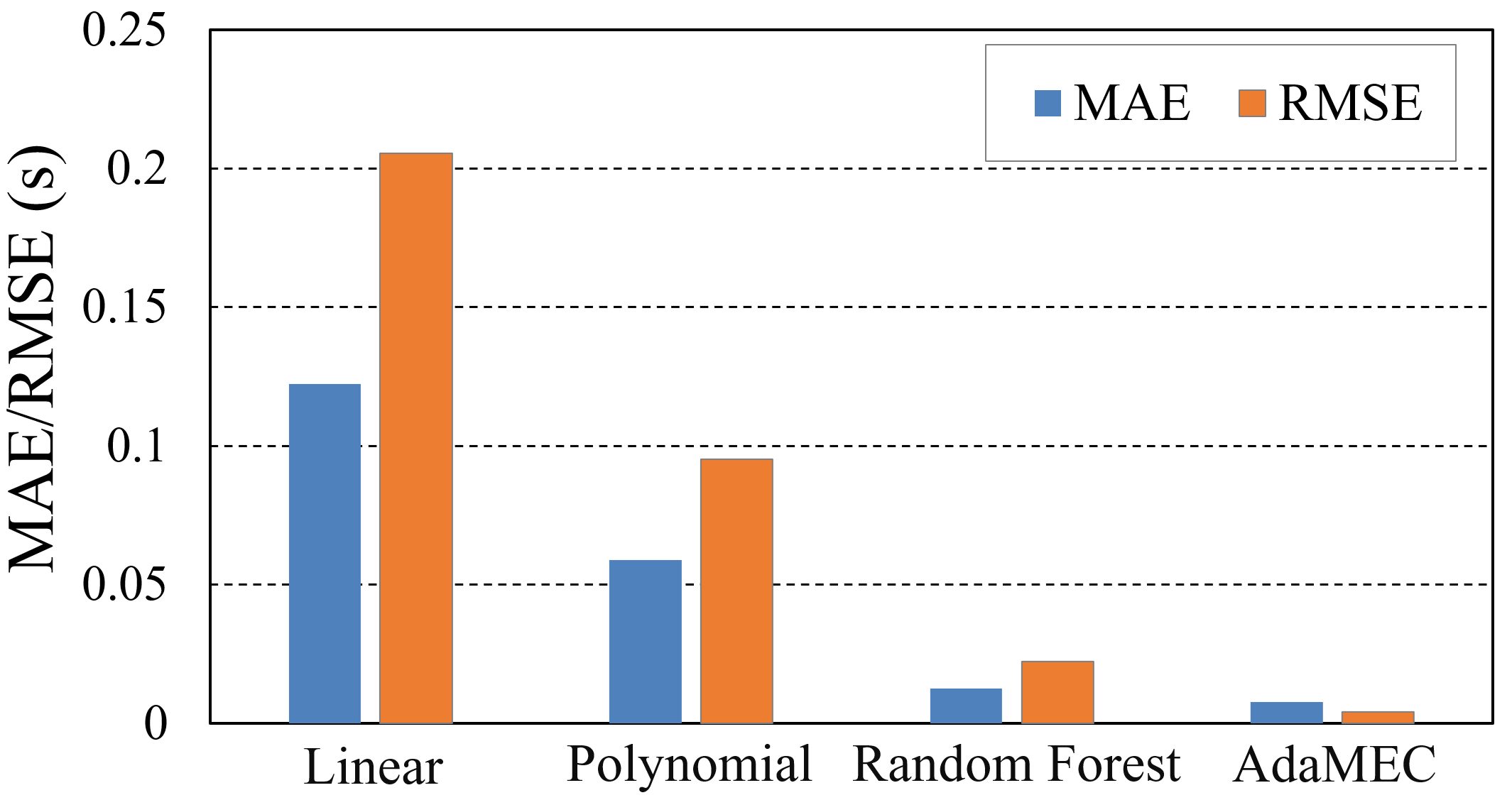}
  \caption{The performance of \systemname's latency predictor on conv, compared with the linear regression-based method, polynomial fitting-based method, and random forest-based method.}
  \label{latency_predictor_performace}
\end{figure} 

\subsubsection{Performance Comparision with baselines} 
We first describe the performance advantages of the proposed latency predictor with three baselines, measured with the mean absolute error (MAE), and root mean squared error (RMSE). 
We design a k-fold cross-validation experiment based on samples of conv, which contributes mainly to the execution latency.
Compared with other baselines, \systemname achieves the lowest prediction error rate, as shown in Fig.~\ref{latency_predictor_performace}. 
In particular, our predictor outperforms the random forest-based (7.8ms MAE and 4.2ms RMSE compared with 12.5ms MAE, 22.4ms RMSE), because the adaptively-enhanced data sampling mechanism captures the non-linear latency patterns between DNN configuration parameters and the execution latency.


\subsubsection{Prediction performance over Different DNNs} 
To verify the universality of \systemname's latency predictor under complex DNN structures, we evaluate it on six typical DNNs (1,200 AlexNets, 2,347 VGGs, 4274 GoogLeNets, 141 ResNets, 240 MobileNets, and 143 Tiny-YOLOs) with five DNN operators. The selected basic DNN operators are shown in Table.~\ref{Sample_space}. 
In addition to MAE and RMSE, we also choose the training score, test score, $\pm$5$\%$ ACC, and $\pm$10$\%$ ACC to evaluate the effectiveness of our latency predictor. Train/test score, with values between 0 and 1, is the coefficient of determination $R^2$ of prediction results, which ignores the influence of the independent variable's magnitude and reflects the ability of the predictor to fit the data. 
The $\pm$5$\%$ /$\pm$10$\%$ ACC indicates the probability that the prediction latency is within the $\pm$5$\%$/$\pm$10$\%$ error boundary. 

As shown in Table.~\ref{tb:prediction_for_models}, \systemname's latency predictor shows strong performance (small MAE and RMSE with high score and accuracy) and generalizability across various models.
Significantly, $>$88.79$\%$ of operators can be predicted within $\pm$10$\%$ error boundary. 
For typical DNNs, we can predict 80.57$\%$ and 94.35$\%$ models within $\pm$5$\%$ and $\pm$10$\%$ error boundaries, respectively.

\textbf{Summary.} \systemnameposs latency predictor achieves a notably high $\pm$10$\%$ accuracy for GoogLeNets, VGGs, and Tiny-YOLOs (\ie 99.89$\%$, 98.65$\%$, and 97.22$\%$).

\subsubsection{Performance Analysis under Dynamic Contexts}
\begin{figure*}[t]
  \centering
  \subfloat[AlexNet.]{
  \includegraphics[height=0.3\textwidth]{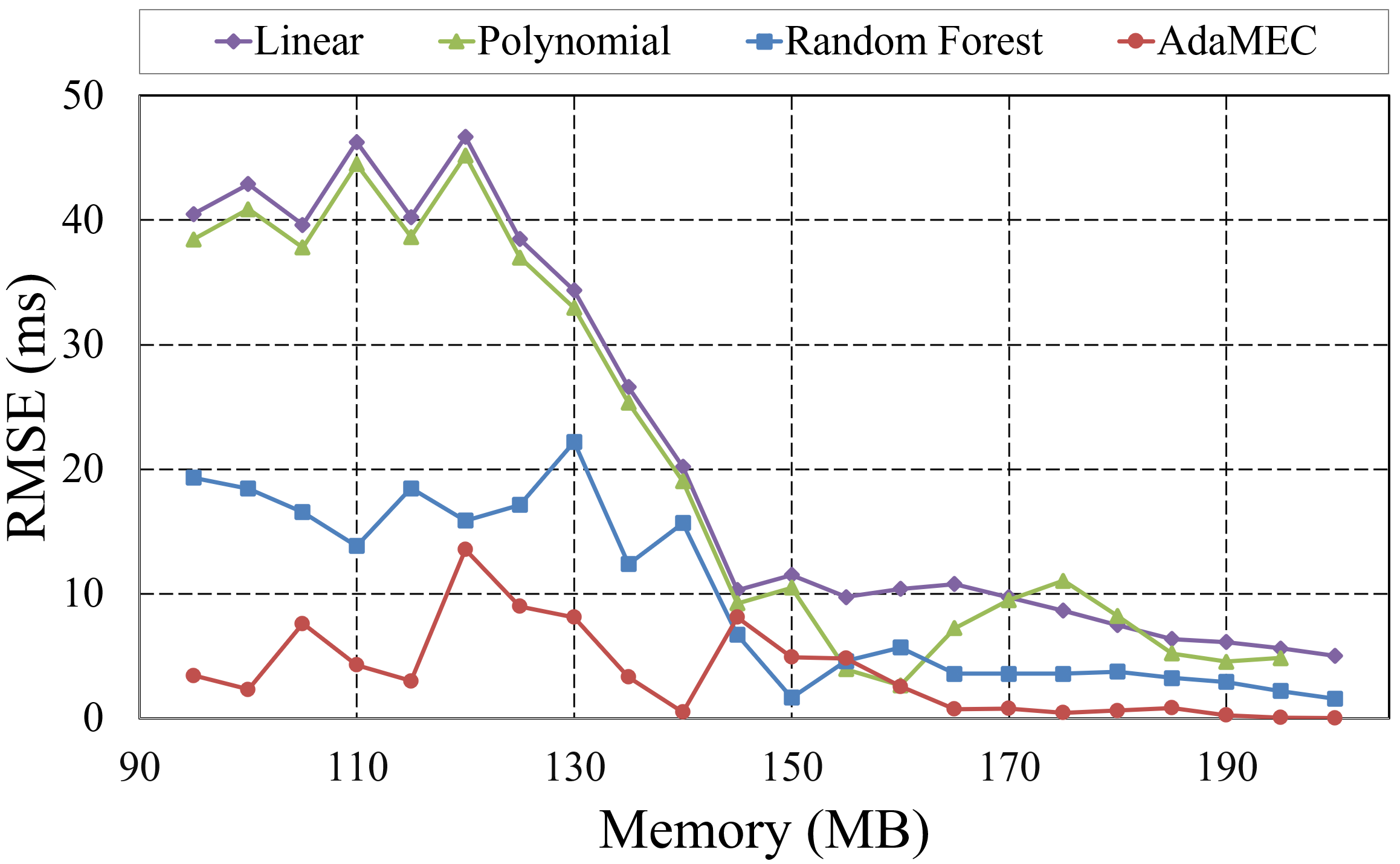}}
  \subfloat[GoogLeNet.]{
  \includegraphics[height=0.3\textwidth]{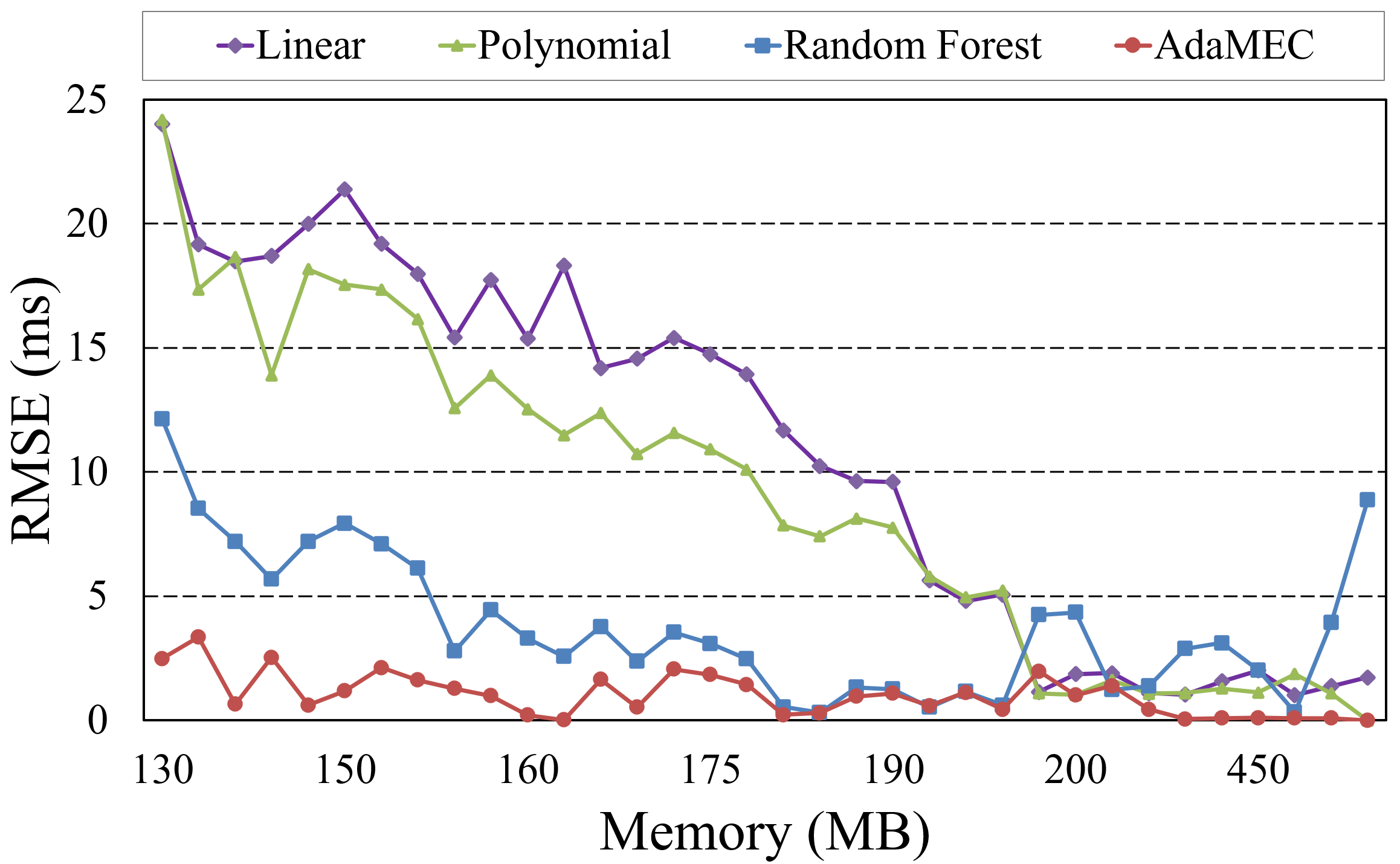}}
\caption{Performance of \systemname's predictor and baselines under dynamic memory budgets.
}
\label{fig_exp_devices}
\vspace{-2mm}
\end{figure*}

\begin{table*}[]
\footnotesize
\centering
\caption{The performance of latency prediction for basic operators and typical variants of DNNs.}
\begin{tabular}{|c|c|c|c|c|c|c|}
\hline
\multicolumn{2}{|c|}{\textbf{Evaluation model}} & \textbf{Train/Test score} & \textbf{MAE (ms)} & \textbf{RMSE (ms)} & \textbf{$\pm$5$\%$ Acc} & \textbf{$\pm$10$\%$ Acc} \\ \hline
\multirow{5}{*}{\textbf{Basic layer type}} & Conv & 0.96/0.87 & 7.80 & 4.20 & 82.08$\%$ & 97.08$\%$ \\ \cline{2-7} 
 & FC & 0.91/0.86 & 0.05 & 0.06 & 74.45$\%$ & 90.01$\%$ \\ \cline{2-7} 
 & BN & 0.94/0.85 & 4.76 & 7.18 & 70.69$\%$ & 88.79$\%$ \\ \cline{2-7} 
 & Maxpool & 0.96/0.68 & 64.90 & 125.00 & 79.72$\%$ & 97.36$\%$ \\ \cline{2-7} 
 & Avgpool & 0.97/0.86 & 4.84 & 10.10 & 79.17$\%$ & 90.00$\%$ \\ \hline
\multirow{6}{*}{\textbf{Typical model type}} & AlexNets & 0.94/0.95 & 1.79 & 2.42 & 80.55$\%$ & 96.34$\%$ \\ \cline{2-7} 
 & VGGs & 0.99/0.94 & 5.03 & 10.71 & 93.83$\%$ & 98.65$\%$ \\ \cline{2-7} 
 & GoogLeNets & 0.96/0.96 & 71.97 & 128.1 & 96.82$\%$ & 99.89$\%$ \\ \cline{2-7} 
 & ResNets & 0.90/0.89 & 105.00 & 253.67 & 68.33$\%$ & 80.56$\%$ \\ \cline{2-7} 
 & Tiny-YOLOs & 0.96/0.94 & 6.53 & 9.64 & 72.22$\%$ & 97.22$\%$ \\ \cline{2-7} 
 & MobileNets & 0.99/0.97 & 9.11 & 13.63 & 71.67$\%$ & 93.33$\%$ \\ \hline
\end{tabular}
\label{tb:prediction_for_models}
\end{table*}

We further conduct experiments to evaluate the performance stability of \systemname's latency predictor under dynamic deployment contexts. 
Specifically, we use Docker on Raspberry Pi 4B to simulate the change of memory resource budgets during running. AlexNet and GoogleNets are selected to execute the image recognition application.
As shown in Fig.~\ref{fig_exp_devices}, compared with baselines, \systemname's latency predictor consistently achieves accurate latency prediction under dynamic memory resources. 
When available memory resources are insufficient, the prediction accuracy of baselines is significantly reduced, while \systemname predicts the latency with 0.49$\sim$8.09ms RMSE for AlexNet and 0.07$\sim$3.36ms RMSE for GoogLeNet under different memory resources, which is significantly superior and stable compared to baselines. 

\textbf{Summary.} \systemname's latency predictor can achieve accurate latency predictions for various and dynamic memory resource budgets, providing the foundation for context-adaptive and efficient DNN partition and offloading.

\section{Related Work}
\label{sec:related}
\subsection{DNN Partition for Mobile Edge Computing}


DNN-powered intelligent applications and services (\eg personal recommendations \cite{zhao2020tkddrecommention}, live video analytics \cite{hung2018videoedge}, companion robotics \cite{saunders2015robotics}) have become a prospective trend.
However, DNNs are typically computationally and storage-intensive, hindering their utilization on resource-constrained mobile and embedded devices (\eg smartphones, wearables, IoT devices). To this end, DNN partition techniques \cite{zhou2019edge, wang2020convergence} are presented to partition DNNs into several parts and distribute different parts to multiple mobile devices for collaborative computing.

The DNN partition methods include two main technical routes, serial computation offloading \cite{kang2017neurosurgeon, eshratifar2019jointdnn,laskaridis2020spinn,zhang2020towards} and parallel computation offloading \cite{zhao2018deepthings,teerapittayanon2017distributed,mao2017modnn,huang2020clio}.
The serialized offloading takes the model layer as the smallest unit to partition DNNs. For example, Neurosurgeon \cite{kang2017neurosurgeon} creates a lightweight schedular to find the optimal partition point for DNNs with chain structure.
QDMP \cite{zhang2020towards} focuses on the partitioning of the advanced DNNs with DAG structure, and formulates it as a min-cut problem to find the appropriate partition.
Similarly, by modeling the task-specific module as a DAG, Distream \cite{zeng2020distream} designs the partitioner to balance the workloads between smart cameras and edge clusters. 
CAS \cite{wang2021context} adopts a heuristic runtime search algorithm to accelerate the decision process for finding an optimal partition in dynamic contexts. 
The parallel offloading method requires a fine-grained design inside the layer to obtain modules that can be computed in parallel. 
In Deepthings \cite{zhao2018deepthings}, DNN is partitioned into multiple independent tasks at the feature map level, to enable parallel execution on multiple devices.
DDNN \cite{teerapittayanon2017distributed} studies how to reduce the inference latency in distributed DNNs, including the cloud, edge, and distributed end devices.
Clio \cite{huang2020clio} partitions output feature maps into different slices and adaptively sends them to the cloud for execution. 


However, most of them ignore practical difficulties encountered in the real-world deployment of DNNs and assume that the entire DNNs have been deployed on devices in advance, which is unacceptable for resource-constrained edge devices and leads to tight coupling between DNN partition and edge deployment. 
Once the deployment context changes (\eg the resources availability, network conditions, and latency requirements), the whole process, including DNN re-partition, storage, re-transmission, and re-offloading of partitioned modules, will be affected and need to re-run for adaptation to the new running context. This process results in significant unnecessary storage and computing resource consumption, which reduces the efficiency of collaborative computing.

\systemname introduces the novel \textit{once-for-all DNN pre-partition} and \textit{context-adaptive DNN atom combination and offloading} approaches to guide the fine-grained atom-based DNN partition and efficient computation offloading.

\subsection{Latency-aware DNN Partition Execution}
Many mobile applications (\eg face recognition~\cite{doukas2010face}, health monitoring~\cite{kavitha2021iot}, and speech recognition~\cite{he2019speech}) are interactive and be sensitive to inference latency.
However, measuring the inference latency by directly deploying DNN models on physical platforms is laborious and expensive due to the wide variety of model structures and platform resources. 
Consequently, researchers are increasingly committed to building latency predictors rather than measuring the inference latency on mobile and embedded devices.

Existing works have investigated the latency prediction of DNN inference from different directions.
First, the FLOPs-based methods~\cite{liu2018darts, tan2019mnasnet, yu2021auto} predict the DNN inference latency based on the floating-point operations per second (FLOPs) or multiply-accumulate operations (MACs).
Second, the Graph Convolutional Networks (GCN)-based methods~\cite{dudziak2020brp} embed DNN architectures into graph structures and adopt GCN to capture the relationship between execution paths in the graph and the inference latency. 
Third, the operator-based methods~\cite{chen2018learning, mendis2019ithemal, adams2019learning} build the latency lookup table or regression model for model operators and sum up the latency of all operators in an estimation model to obtain the overall inference latency.
Despite their progress, it is non-trivial to provide a precise latency predictor. It depends on the model types (\eg convolutional neural networks (CNN), recurrent neural networks (RNN)), model hyperparameters (\eg kernel size, input channel), operation contexts (\eg the memory supply), and other factors.
Also, the above efforts target the single-device scenario, ignoring the distributed characteristics of collaborative inference with DNN partition on multiple edge devices.




To predict the latency of distributed DNN execution, Neurosurgeon~\cite{kang2017neurosurgeon} creates a latency predictor for every possible model partition point between mobile devices and the cloud by using linear regression.
%
Edgent~\cite{li2019edge} builds a configuration map constructor using the regression method.
However, these methods based on the offline regression can only retain limited static information (\eg model structures), and it is difficult to deal with the \textit{dynamic nature} of network conditions and distributed platform resources~\cite{jeong2018ionn, yao2021context}.

\systemname enables a \textit{model-independent} and \textit{resource-independent} latency predictor for guiding the DNN partition decision in the constantly changing execution context of distributed mobile and edge devices.
Specifically, it formulates the latency at a fine-grained level (\eg kernel size, input channels, memory of devices).

\section{Conclusion}
\label{sec:conclusion}

This paper presents \systemname, a context-adaptive and dynamically-combinable DNN deployment framework for mobile edge computing. 
It decouples the end-to-end process of distributed DNN deployment over mobile and edge devices into two independent steps, \ie once-for-all DNN pre-partition, and context-adaptive DNN atom offloading.
In addition, it presents a latency predictor to provide timely and accurate latency feedback for assessing candidate DNN deployment strategies. 
Evaluation under six typical DNNs across three mobile and edge devices indicates the advantages of \systemname in terms of inference latency reduction by up to 62.14\%, and memory reduction on both mobile and edge devices by up to 55.21\%.
In the future, we plan to extend our framework to integrate diverse tasks and consider the impact of comprehensive deployment contexts for mobile edge computing.


\bibliographystyle{ACM-Reference-Format}
\bibliography{sample-base}

\end{document}